\documentclass[twocolumn]{aastex631}

\usepackage{appendix}
\usepackage{mathrsfs}
\usepackage{bold-extra}

\usepackage{lineno}

\begin{document}

\title{Chasing the beginning of reionization in the JWST era}

\correspondingauthor{Christopher Cain}
\email{clcain3@asu.edu}

\author[0000-0001-9420-7384]{Christopher Cain}
\affiliation{School of Earth and Space Exploration, Arizona State University,
Tempe, AZ 85287-6004, USA}

\author{Garett Lopez}
\affiliation{Department of Physics and Astronomy, University of California, Riverside, CA 92521, USA}

\author{Anson D'Aloisio}
\affiliation{Department of Physics and Astronomy, University of California, Riverside, CA 92521, USA}

\author[0000-0002-8984-0465]{Julian B.~Mu\~noz}
\affiliation{Department of Astronomy, The University of Texas at Austin, 2515 Speedway, Stop C1400, Austin, TX 78712, USA}

\author[0000-0003-1268-5230]{Rolf A.~Jansen}
\affiliation{School of Earth and Space Exploration, Arizona State University,
Tempe, AZ 85287-6004, USA}

\author[0000-0001-8156-6281]{Rogier A. Windhorst} %%% Rogier.Windhorst@gmail.com
\affiliation{School of Earth and Space Exploration, Arizona State University,
Tempe, AZ 85287-6004, USA}

\author{Nakul Gangolli}
\affiliation{Department of Physics and Astronomy, University of California, Riverside, CA 92521, USA}

\begin{abstract}

Recent JWST observations at $z > 6$ may imply galactic ionizing photon production above prior expectations.  Under observationally motivated assumptions about escape fractions, these suggest a $z \sim 8-9$ end to reionization, in tension with the $z <  6$ end required by the Ly$\alpha$ forest.  In this work, we use radiative transfer simulations to understand what different observations tell us about when reionization ended and when it started.  We consider a model that ends too early ($z_{\rm end} \approx 8$) alongside two more realistic scenarios with $z_{\rm end} \approx 5$: one starting late ($z \sim 9$) and another early ($z \sim 13$).  We find that the latter requires up to an order-of-magnitude evolution in galaxy ionizing properties at $6 < z < 12$, perhaps in tension with measurements of $\xi_{\rm ion}$ by JWST, which indicate little evolution.  We study how these models compare to recent measurements of the Ly$\alpha$ forest opacity, mean free path, IGM thermal history, visibility of $z > 8$ Ly$\alpha$ emitters, and the patchy kSZ signal from the CMB.  We find that neither of the late-ending scenarios is strongly disfavored by any single data set. However, a majority of observables, spanning several distinct types of observations, prefer a late start.  Not all probes agree with this conclusion, hinting at a possible lack of concordance arising from deficiencies in observations and/or theoretical modeling.  Observations by multiple experiments (including JWST, Roman, and CMB-S4) in the coming years will establish a concordance picture of reionization's beginning or uncover such deficiencies.  

\end{abstract}


\section{Introduction} \label{sec:intro}

Despite an explosion of new data in the past decade probing cosmic reionization, little is known about how and when the process began.  The ending of reionization, believed to occur at $5 < z < 6$, has been constrained by observations of the Ly$\alpha$ forest of high-redshift QSOs~\citep{Becker2015,Kulkarni2019,Keating2019,Nasir2020,Qin2021,Bosman2021}.  Direct (and indirect) measurements of the mean free path from QSO spectra~\citep{Worseck2014,Becker2021,Zhu2023,Gaikwad2023,Davies2024}, of the IGM photo-ionization rate~\citep{Wyithe2011,Calverley2011,DAloisio2018,Becker2021,Gaikwad2023}, and of the IGM thermal history at $z \geq 5.5$~\citep{Gaikwad2020} have further corroborated this picture.  The electron scattering optical depth to the CMB~\citep[$\tau_{\rm es}$,][]{Planck2018} constrains the midpoint of reionization to be $z \sim 7.5$.  Evidence of damping wings in high-redshift $z \sim 7-8$ QSOs~\citep{Davies2018,Wang2020,Yang2020a} and measurements of the neutral fraction based on (non)detections of Ly$\alpha$ emitters~\citep[LAEs, e.g.][]{Mason2018a,Mason2019,Whitler2020,Jung2020} indicate that the IGM was partially neutral at these redshifts.  

Understanding the timeline of reionization is crucial for revealing the nature of the ionizing sources that drove it.  Explaining how reionization could be driven by galaxies alone has been historically challenging thanks to the high early measurements of $\tau_{\rm es}$~\citep{Dunkley2009,Komatsu2011} which required very early and/or extended reionization histories.  This tension eased as the measured value of $\tau_{\rm es}$ steadily decreased.  Several recent works~\citep[e.g.][]{Bouwens2015,Robertson2015,Finkelstein2019,Matthee2022} have showed that galaxies can complete reionization by $z \approx 6$ under physically reasonable assumptions about their ionizing properties.  Recent evidence for an end later than $z = 6$ further relaxed demands on galaxy ionizing output (although see~\citealt{Davies2021b,Davies2024b}).  Concurrent efforts demonstrated that AGN are unlikely to have contributed the majority of the ionizing budget responsible for reionization~(e.g. \citealt{Dayal2020,Trebitsch2023}, although see~\citealt{Madau2015,Madau2024}).  

These findings paint a relatively simple, consistent picture of reionization: it ended at $5 < z < 6$, was in progress at $z \sim 7-8$, and was likely driven by galaxies.  Prior to JWST, the precise timing of reionization's midpoint and especially its early stages were not tightly constrained.  Constraints on reionization's midpoint from $\tau_{\rm es}$~\citep{Planck2018} spanned a range of $\pm 0.75$ in redshift (at $1\sigma$), and few direct constraints on the first half of reionization existed.  The space of models proposed by the aforementioned works span a wide range of possibilities\footnote{The scenarios proposed by~\citealt{Finkelstein2019} (early-starting, gradually ending reionization), and~\citealt{Matthee2022} (late-starting, rapidly ending) reionization, roughly bracket the proposed possibilities.  } without contradicting observations.  Indeed, one important goal for JWST is to probe the properties of galaxies at $z > 7-8$ in hopes of learning more about reionization's early stages. 

\begin{figure}
    \centering
    \includegraphics[scale=0.18]{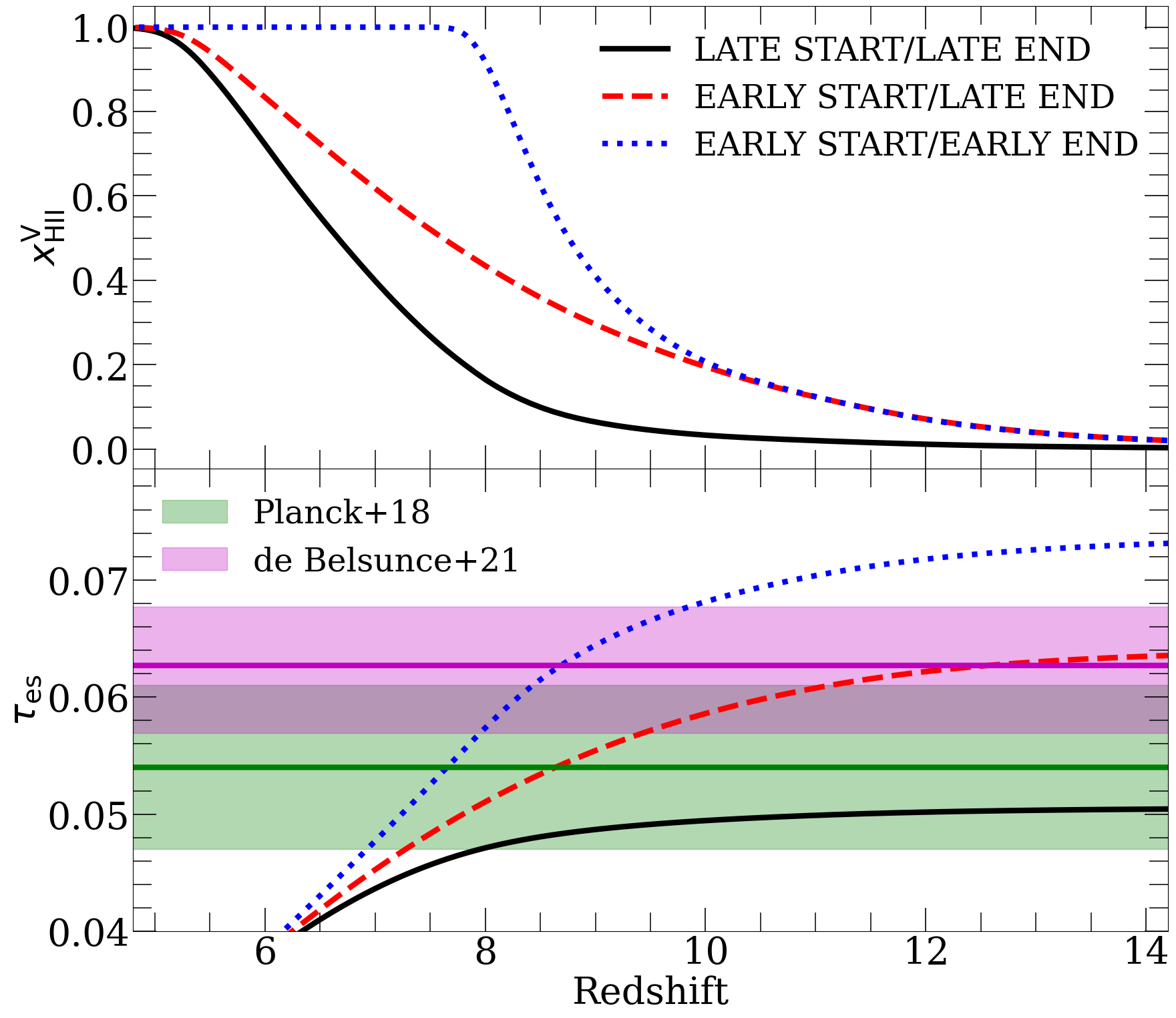}
    \caption{Summary of the reionization scenarios studied in this work.  Broadly speaking, they represent the three types of possible reionization histories.  Top: volume-averaged ionized fraction $x_{\rm  HII}^{\rm V}$ vs. redshift for the \textsc{late start/late end}, \textsc{early start/late end}, and \textsc{early start/early end} models.  These models have $z_{\rm end} \approx 5$, $5$, $8$ and $z_{\rm mid} \approx 6.5$, $7.5$, $8.5$, respectively.  Bottom: CMB electron scattering optical depth $\tau_{\rm es}$ for each model, compared to the results of~\citealt{Planck2018,deBelsunce2021} (shaded regions denote $\pm 1\sigma$).  The \textsc{late start/late end} model is within $1\sigma$ of Planck, while \textsc{early start/late end} is within $1 \sigma$ of the re-analysis by~\citealt{deBelsunce2021}.  }
    \label{fig:ion_history_tau}
\end{figure}

However, the first JWST results may be complicating, as much as clarifying, our understanding of reionization.  JWST has allowed for measurements of the UV luminosity function (UVLF) at much higher redshifts than HST, up to $z \sim 14$~\citep{Finkelstein2024,Adams2024,Donnan2024}.  It has also allowed us to measure the ionizing efficiency of galaxies, $\xi_{\rm ion}$, above $z = 6$~\citep{Endsley2023,Simmonds2024,Pahl2024}.  Recently,~\citealt{Munoz2024} pointed out that a face-value interpretation of recent UVLF and $\xi_{\rm ion}$ measurements from JWST (\citealt{Donnan2024,Simmonds2024}) combined with observationally motivated assumptions about ionizing escape fractions ($f_{\rm esc}$) suggests reionization ended around $z \sim 8-9$, inconsistent at $> 2\sigma$ with the Planck $\tau_{\rm es}$ measurement\footnote{Although the tension is slightly smaller with the recent re-measurement of $\tau_{\rm es}$ from Planck data by~\citealt{deBelsunce2021} - see Figure~\ref{fig:ion_history_tau}.  }.  This result represents a stark reversal from the historical problem of galaxies producing too few ionizing photons to complete reionization on time~\citep[see ][and references therein]{Robertson2015}.  Several bright Ly$\alpha$ emitters (LAEs) at $z > 8$~\citep{Zitrin2015,Larson2022,Bunker2023,Curti2024,Tang2024b} have also been observed, with the highest-redshift detection to date at $z = 10.6$~\citep{Bunker2023} by JWST.  These observations may be surprising if the IGM is mostly neutral at these redshifts, since observing Ly$\alpha$ emission requires some level of ionization around galaxies~\citep{Mason2020}.  

The top panel of Figure~\ref{fig:ion_history_tau} shows the volume-averaged ionized fraction ($x_{\rm HII}^{\rm V}$) for the three reionization models we will study in this work.  The \textsc{early start/early end} model is motivated by the aforementioned findings of~\citealt{Munoz2024}, and has a midpoint (endpoint) of $z_{\rm mid} = 8.5$ ($z_{\rm end} = 8$).  The \textsc{late start/late end} model is motivated by Ly$\alpha$ forest observations at $5 < z < 6$ and the Planck $\tau_{\rm es}$ measurement, has $z_{\rm mid} = 6.5$ and $z_{\rm end} = 5$.  The \textsc{early start/late end} model also has $z_{\rm end} = 5$, but an earlier midpoint ($z_{\rm mid} = 7.5$) and a start at $z \sim 13$.  These three models broadly represent the three types, or categories, of possible reionization histories\footnote{Note that we do not include a hypothetical \textsc{late start/early end} model here, since it would be essentially an instantaneous reionization history and, as we will see, would not add meaningfully to the conclusions in this work. 
 }.  The bottom panel shows $\tau_{\rm es}$ for each, compared with the~\citealt{Planck2018} measurement (black) and the recent re-analysis of Planck data from~\citealt{deBelsunce2021}, with shaded regions denoting $\pm 1\sigma$ uncertainties.  The \textsc{late start/late end} model is consistent within $1 \sigma$ with~\citealt{Planck2018}, and the \textsc{early start/late end} case is similarly consistent with~\citealt{deBelsunce2021}\footnote{It is noted in~\citet{deBelsunce2021} that their measurement of $\tau_{\rm es}$ is slightly higher than that of~\citet{Planck2018} because they treat temperature and polarization jointly in their analysis.  }.  

In this work, we study the observational properties of the three reionization models shown in Figure~\ref{fig:ion_history_tau} using radiative transfer (RT) simulations.  We will demonstrate, in accord with previous works, that observations from the $z \lesssim 6$ Ly$\alpha$ forest strongly disfavor the \textsc{early start/early end} model, in agreement with $\tau_{\rm es}$.  We will also compare the two late-ending models to a wide range of observations with the goal of understanding whether existing data from multiple sources points in the direction of a late or an early start to reionization.  This work is organized as follows. \S\ref{sec:JWST} describes how we calibrate our three reionization models and discusses their implications for galaxy properties in light of recent JWST observations.  In \S\ref{sec:numerical}, we describe our methods for running RT simulations of reionization and forward-modeling various observables.  We compare our models to several sets of complementary observations in \S\ref{sec:results}, discuss the implications of our findings in \S\ref{sec:disc}, and conclude in \S\ref{sec:conc}.  After discussing an observable, we will bold-face the name of the reionization model (if any) preferred by that observable.  Throughout, we assume the following cosmological parameters: $\Omega_m = 0.305$, $\Omega_{\Lambda} = 1 - \Omega_m$, $\Omega_b = 0.048$, $h = 0.68$, $n_s = 0.9667$ and $\sigma_8 = 0.82$, consistent with~\citealt{Planck2018} results. All distances are in co-moving units unless otherwise specified.  

\section{Implications of JWST galaxy observations} \label{sec:JWST}

\subsection{Reionization Model Calibration}
\label{subsec:models}

The main input to our RT simulations (described in \S\ref{sec:numerical}) is the globally averaged ionizing photon emissivity of sources verses redshift, $\dot{N}_{\gamma}(z)$.  In this section we describe how we use a combination of JWST observations and Ly$\alpha$ forest data to construct (or ``calibrate'') $\dot{N}_{\gamma}(z)$ for our three models.  

\begin{figure*}
    \centering
    \includegraphics[scale=0.33]{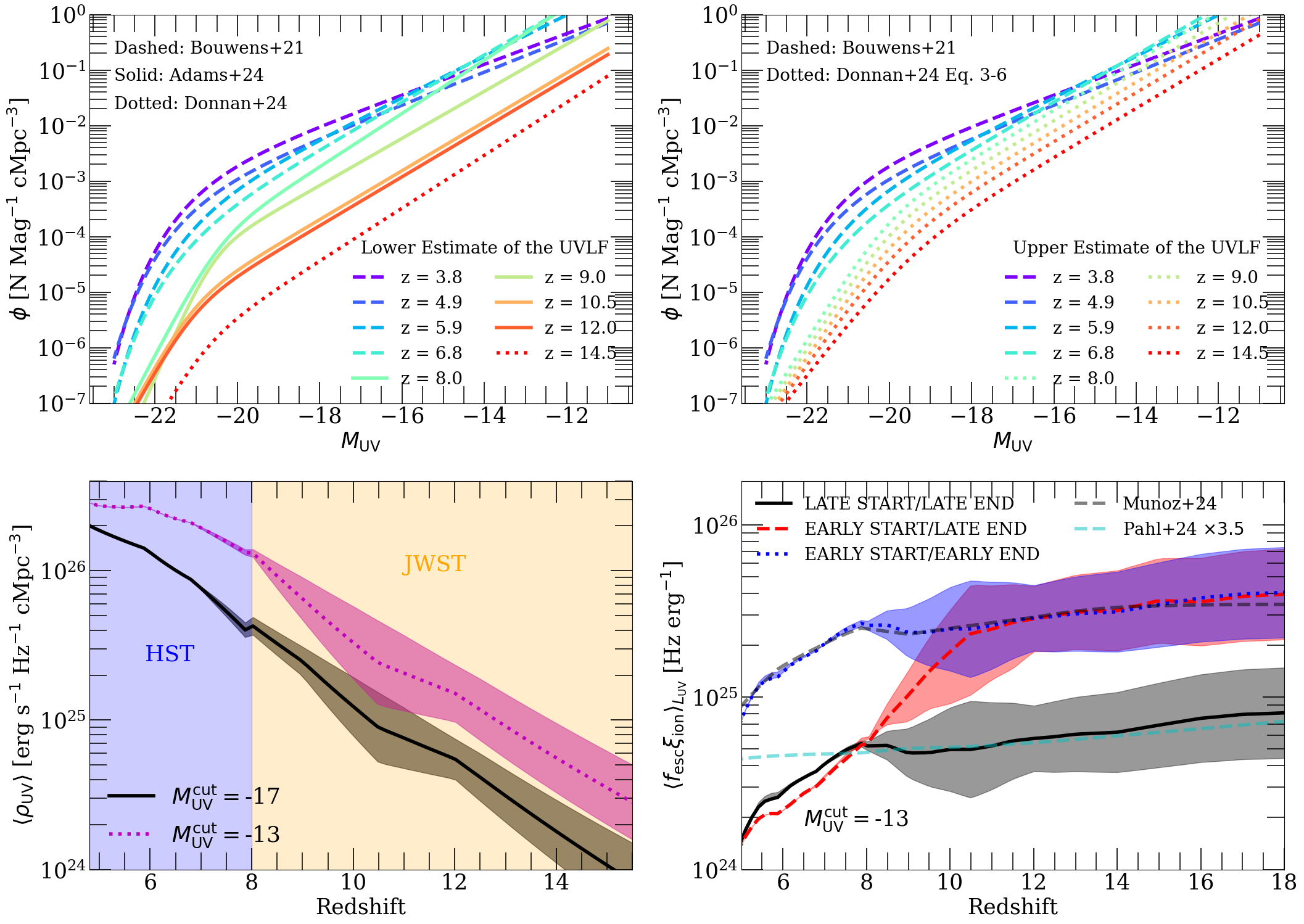}
    \caption{Evolution of galaxy properties and calibration of $\dot{N}_{\gamma}(z)$ for our reionization scenarios.  {\bf Top Left:} measured UVLFs from \citealt{Bouwens2021} at $z < 8$, \citealt{Adams2024} at $8 < z < 12.5$, and \citealt{Donnan2024} at $z = 14.5$.  {\bf Top Right:} The same, but using UVLF parameters from Eq. 3-6 of~\citealt{Donnan2024} at $z \geq 8$.  These two sets are different by a factor of $\sim 3$ at $z > 8$, roughly bracketing observational uncertainty in the UVLF.  {\bf Lower left:} integrated $\rho_{\rm UV}(z)$ for cutoff magnitudes $M_{\rm UV}^{\rm cut} = -17$ and $-13$.  The shaded regions denote the spread between the two sets of UVLFs in the top panels, and the curves denote the logarithmic averages.  {\bf Lower Right:} product of $f_{\rm esc}$ and $\xi_{\rm ion}$, weighted by $L_{\rm UV}$ and averaged over the galaxy population (Eq.~\ref{eq:fesc_xion}).  The colored lines show this quantity for our models, and the shaded regions show the spread resulting from observational uncertainty in the UVLF (at fixed $\dot{N}_{\gamma}(z)$).  The faded gray dashed curve shows one of the observationally motivated models assumed in~\citealt{Munoz2024} (with $M_{\rm UV}^{\rm cut} = -13$) and the faded cyan dashed curve shows the model of~\citealt{Pahl2024} scaled up by a factor of $3.5$.  Our \textsc{early start/early end} scenario (by construction) approximately matches the~\citealt{Munoz2024} model.  The \textsc{late start/late end} case shares the same redshift evolution, but is scaled down by a factor of $5$ to recover agreement with the $z \leq 6$ Ly$\alpha$ forest.  The \textsc{early start/late end} model is calibrated to agree with the~\citealt{Munoz2024} result at $z \geq 10$, but it also required to agree with the Ly$\alpha$ forest.  This model demands a factor of $\approx 10$ decline in $\langle f_{\rm esc} \xi_{\rm ion}\rangle_{L_{\rm UV}}$ between $z = 12$ and $6$ - much steeper evolution than suggested by the~\citealt{Munoz2024} or~\citealt{Pahl2024} findings.  }
    \label{fig:UVLF_plot}
\end{figure*}

Our starting point is the amount of non-ionizing UV light produced by galaxies, which is quantified by the UV luminosity function (UVLF).  This has been measured up to $z \sim 14$ by JWST using both photometry and spectroscopy~\citep[e.g.][]{Harikane2024,Finkelstein2024,Adams2024,Donnan2024}.  The top two panels of Figure~\ref{fig:UVLF_plot} show two sets of UVLFs that we use in our analysis.  In both panels, the dashed curves show the $z < 8$ UVLFs measured by~\citealt{Bouwens2021} with HST. In the left panel, the solid curves denote the double-power-law (DPL) fits from~\citealt{Adams2024} at $8 \leq z \leq 12$, and the dotted curve is the measurement from~\citealt{Donnan2024} at $z = 14.5$.  In the right panel, the dotted curves show results for the redshift-dependent UVLF parameters given in Eq. 3-6 of~\citealt{Donnan2024} (which is a best-fit to measurements from~\citealt{Bowler2016,Bowler2020,Donnan2023,McLeod2023}).  At $z > 8$, the UVLFs in the left panel are a factor of $\sim 3$ lower than those in the right panel.  We will use these sets to roughly bracket observational uncertainties on the UVLF.  

The lower left panel shows the integrated UV luminosity density, $\rho_{\rm UV}$, vs. redshift.  The curves show the logarithmic average of $\rho_{\rm UV}$ calculated using the two sets of UVLFs in the top panels, and the shaded regions show the spread between them.  For illustration, we integrate the UVLF down to two limiting magnitudes - a bright cutoff of $M_{\rm UV}^{\rm cut} = -17$ (black) and a fainter $M_{\rm UV}^{\rm cut} = -13$ (magenta).  The background shading denotes the redshift ranges covered by HST and JWST data.  Note that at $z > 8$, the redshift evolution of $\rho_{\rm UV}(z)$ is fairly insensitive to $M_{\rm UV}^{\rm cut}$, the main difference being normalization.  This is because both sets of UVLFs have faint end slopes close to $\alpha = -2.1$ at $z \geq 8$, with little evolution across that redshift range~\citep{Kravtsov2024}.  This is in contrast to some pre-JWST expectations, which predicted a steepening of the faint-end slope with redshift and a corresponding shallow evolution in $\rho_{\rm UV}$ for faint $M_{\rm UV}^{\rm cut}$~\citep[e.g.][]{Finkelstein2019}.  Instead, $\rho_{\rm UV}$ evolves quickly with redshift, with a factor of $\sim 10$ evolution between $z = 8$ and $13$.  

The lower right panel quantifies the redshift evolution of galaxy ionizing properties in our models using
\begin{equation}
    \label{eq:fesc_xion}
    \langle f_{\rm esc} \xi_{\rm ion} \rangle_{L_{\rm UV}} \equiv \frac{\dot{N}_{\gamma}(z)}{\rho_{\rm UV}(z)}
\end{equation}
This is the UV-luminosity ($L_{\rm UV}$)-weighted average of the product of $f_{\rm esc}$ and $\xi_{\rm ion}$ for the galaxy population.  The black, red, and blue curves show this quantity for our three reionization models (see legend) assuming the average $\rho_{\rm UV}$ curve for $M_{\rm UV}^{\rm cut} = -13$ (lower left panel).  The shaded regions show the spread in this quantity (for fixed $\dot{N}_{\gamma}(z)$) arising from observational uncertainty in $\rho_{\rm UV}(z)$, as shown in the lower left.  We calibrate $\dot{N}_{\gamma}(z)$ for the \textsc{early start/early end} model such that its $\langle f_{\rm esc} \xi_{\rm ion} \rangle_{L_{\rm UV}}$ agrees with the observationally motivated model from~\citealt{Munoz2024} (which assumes $M_{\rm UV}^{\rm cut} = -13$), shown by the faded gray dashed curve.  Consistent with the findings of~\citealt{Munoz2024}, this model ends reionization early at $z \approx 8$ (see  Figure~\ref{fig:ion_history_tau}).  

Our \textsc{late start/late end} scenario is motivated by the possibility that the tension with $\tau_{\rm es}$ and the Ly$\alpha$ forest in the \textsc{early start/early end} model can be solved with a simple redshift-independent re-scaling of $\dot{N}_{\gamma}(z)$.  This could be the case if the measurements of $\xi_{\rm ion}$ assumed in~\citealt{Munoz2024} (from~\citealt{Simmonds2024}) are systematically biased high, and/or if the same is true of the $f_{\rm esc}$ values they inferred from the results of~\citealt{Chisholm2022,Zhao2024}.  We find that scaling $\dot{N}_{\gamma}(z)$ down by a factor of $5$ from the \textsc{early start/early end} case brings the end of reionization to $z \sim 5$.  We then further adjust $\dot{N}_{\gamma}(z)$ at $z < 7$ at the few-percent level until we achieve good agreement with the Ly$\alpha$ forest at $z \leq 6$ (as we will show in \S\ref{subsubsec:lya}).  The cyan-dashed curve shows the observationally motivated model from~\citealt{Pahl2024}, multiplied by $3.5$, which also has redshift evolution similar to this scenario.   

However, the \textsc{late start/late end} scenario is not the only possibility allowed by $\tau_{\rm es}$ and the Ly$\alpha$ forest.  It could also be that $\dot{N}_{\gamma}(z)$ is consistent with the~\citealt{Munoz2024} model at the highest redshifts ($z \gtrsim 10$), but declines at lower redshifts in such a way that reionization ends at $z < 6$, as required by the Ly$\alpha$ forest.  This scenario is represented by our \textsc{early start/late end} model (red dashed curve).  This model also ends reionization at $z = 5$, and by adjusting its $\dot{N}_{\gamma}(z)$ at $z < 10$, we can also achieve agreement with the Ly$\alpha$ forest.  This model requires a factor of $\approx 10$ decline in $\langle f_{\rm esc} \xi_{\rm ion} \rangle_{L_{\rm UV}}$ between $z = 10$ and $6$, a decrease much steeper than suggested by the results of of~\citealt{Munoz2024} and~\citealt{Pahl2024}.  This could be achieved if $\xi_{\rm ion}$, $f_{\rm esc}$, $M_{\rm UV}^{\rm cut}$, or some combination of these evolves significantly across this redshift range (see \S\ref{subsec:scenarios}).  

\subsection{Is the \textsc{early start/late end} model plausible?}
\label{subsec:scenarios}

\begin{figure}
    \centering
    \includegraphics[scale=0.23]{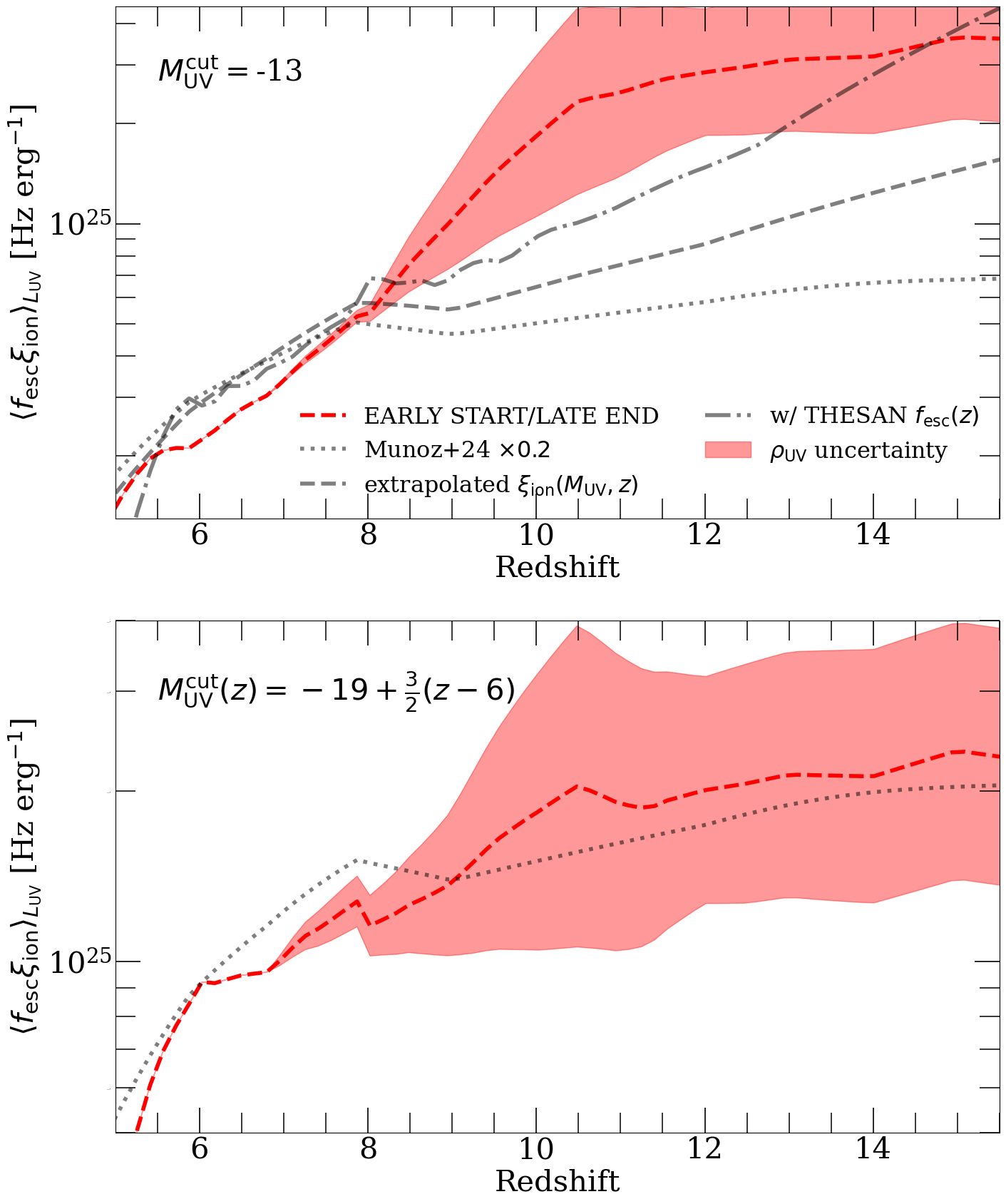}
    \caption{Summary of mechanisms that could allow for the \textsc{early start/late end} model.  {\bf Top:} $\langle f_{\rm esc} \xi_{\rm ion}\rangle_{L_{\rm UV}}$ (as in Figure~\ref{fig:UVLF_plot}), alongside several observationally and theoretically motivated scenarios.   The grey dotted curve shows the~\citealt{Munoz2024} model multiplied by $0.2$, which matches the \textsc{late start/late end model}. 
    The gray dashed curve is the same model, but with caps on extrapolation of $\xi_{\rm ion}$ with $M_{\rm UV}$ and redshift imposed by~\citealt{Munoz2024} removed (see text).  The dot-dashed curve further replaces the model for $f_{\rm esc}$ used in~\citealt{Munoz2024} with the average $f_{\rm esc}(z)$ measured in the flagship THESAN simulation~\citep{Yeh2022}.  This last curve reproduces evolution qualitatively consistent with the \textsc{early start/late end} model.  {\bf Bottom:} the same, but assuming $M_{\rm UV}^{\rm cut}$ evolves linearly from $-19$ at $z = 6$ to $-10$ at $z = 12$ (see annotation).  A trends towards fainter $M_{\rm UV}^{\rm cut}$ at higher $z$ causes $\rho_{\rm UV}$ to decline less steeply, requiring shallower redshift evolution of ionizing properties.  See text for further details.  }
    \label{fig:alternatives}
\end{figure}

Given the discrepancy between the evolution of $\langle f_{\rm esc} \xi_{\rm ion} \rangle_{L_{\rm UV}}$ in the \textsc{early start/late end} model and in the~\citealt{Munoz2024} and~\citealt{Pahl2024} models, it is natural to ask whether this scenario is plausible.  The top panel of Figure~\ref{fig:alternatives} shows $\langle f_{\rm esc} \xi_{\rm ion}\rangle_{L_{\rm UV}}$ for this model, alongside observationally and theoretically motivated scenarios that make various assumptions about the redshift evolution of $\xi_{\rm ion}$ and/or $f_{\rm esc}$.  The dotted curve is the~\citealt{Munoz2024} model scaled down by $0.2$, which agrees with the \textsc{late start/late end} case.  The dashed curve is the same, except that we extrapolate $\xi_{\rm ion}$ (Eq.~4 in~\citealt{Munoz2024}) outside the range of $M_{\rm UV}$ and redshift within which it was fit to data.  For the dot-dashed curve, we further replace the observationally motivated $f_{\rm esc}$ prescription assumed in~\citealt{Munoz2024} with the global $f_{\rm esc}(z)$ from the flagship THESAN simulation~\citep{Yeh2022}.  We have re-scaled each of the gray curves by different constants to bring them as close as possible to the \textsc{early start/late end} model.  We obtain the best agreement for the dot-dashed curve, which boosts the redshift evolution of both $\xi_{\rm ion}$ and $f_{\rm esc}$ relative to~\citealt{Munoz2024}.  This shows that the \textsc{early start/late end} model is plausible given evolution in $f_{\rm esc}$ and/or $\xi_{\rm ion}$, but only under ``favorable'' assumptions about both.    

The bottom panel of Figure~\ref{fig:alternatives} illustrates another mechanism that could work in the direction of making reionization start earlier: evolution in $M_{\rm UV}^{\rm cut}$.  The red curve assumes that the $M_{\rm UV}^{\rm cut}$ evolves from $-10$ at $z = 12$ to $-19$ at $z = 6$ $\left(M_{\rm UV}^{\rm cut}(z) = -19 + \frac{3}{2}(z - 6)\right)$, which causes $\rho_{\rm UV}$ to decline much less rapidly with redshift than it does in the bottom left panel of Figure~\ref{fig:UVLF_plot}.  This allows $\langle f_{\rm esc} \xi_{\rm ion}\rangle_{L_{\rm UV}}$ in the \textsc{early start/late end} scenario to evolve less quickly, in better agreement with the~\citealt{Munoz2024} model (dotted curve).  This type of behavior in the galaxy population could arise from decreasing dust obscuration~\citep{Topping2024} and/or feedback from the IGM reducing or shutting off star formation in low-mass halos at lower redshifts~\citep{Wu2019b,Ocvirk2021}.  The evolution in $M_{\rm UV}^{\rm cut}$ assumed in this illustrative example is extreme, and likely ruled out by existing observations~\citep{Atek2018}, but serves to show how an evolving $M_{\rm UV}^{\rm cut}$ could support the \textsc{early start/late end} model.  

These comparisons suggest that a factor of a few of evolution in each of $\xi_{\rm ion}$ and $f_{\rm esc}$, and perhaps some in $M_{\rm UV}^{\rm cut}$, could explain the \textsc{early start/late end} scenario.  Indeed, there is some observational and theoretical support for the idea that $\xi_{\rm ion}$ could increase with redshift and/or be larger in fainter galaxies~\citep{Atek2023,Cameron2023,Simmonds2024}, However, some works find that $\xi_{\rm ion}$ is closer to constant with $M_{\rm UV}$, and perhaps redshift~\citep[][]{Matthee2022,Pahl2024}.  Unfortunately, $f_{\rm esc}$ at these redshifts cannot be directly measured, and there is no consensus in the literature from either indirect observational estimates~\citep[e.g.][]{Naidu2022,Citro2024} or cosmological simulations~\citep[e.g.][]{Trebitsch2018,Rosdahl2022,Kostyuk2023}.  Thus, galaxy observations alone cannot rule out the \textsc{early start/late end} case, although such a model does require more significant evolution in galaxy ionizing properties than suggested by current observations. As such, we conclude that the \textbf{\textsc{late start/late end}} model seems more likely based on existing data.  

\subsection{Ionizing photon budget}
\label{subsec:ion_budget}

\begin{figure}
    \centering
    \includegraphics[scale=0.29]{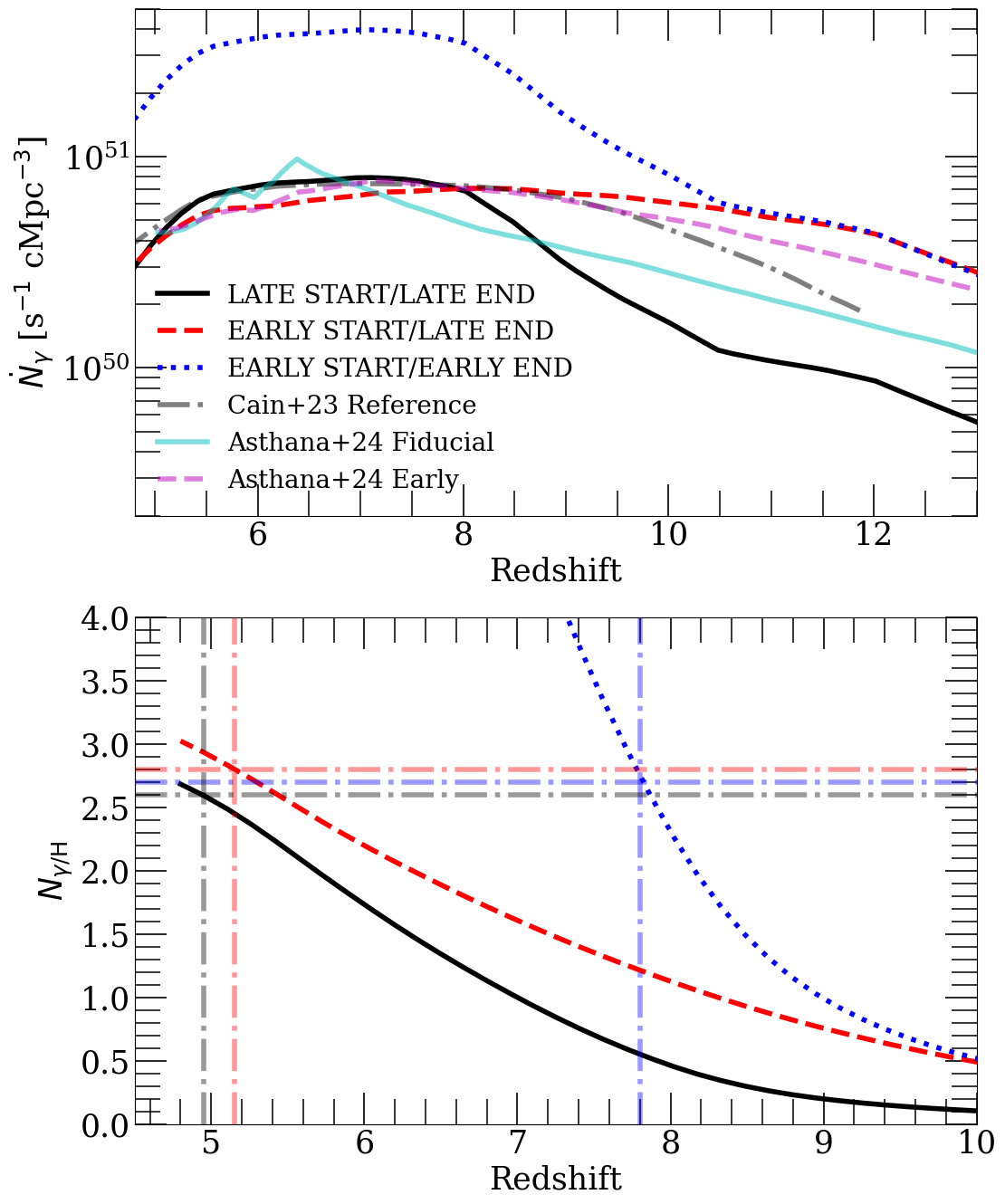}
    \caption{{\bf Top:} ionizing photon emissivity for our models, alongside several others from the literature that are calibrated to match the Ly$\alpha$ forest observations at $z \leq 6$.  Our \textsc{late start/late end} model has the steepest evolution at $z > 9$, and our \textsc{early start/late end} model agrees well with the ``early'' model from~\citealt{Asthana2024}.  {\bf Bottom:} ionizing photon production per H atom by galaxies.  The vertical dashed lines denote the end of reionization in each model, and the horizontal lines indicate the number of photons per H atom required to complete reionization.  }
    \label{fig:emissivities}
\end{figure}

We show the $\dot{N}_{\gamma}(z)$ for our models in the top panel of Figure~\ref{fig:emissivities}, alongside several others from the literature that also match the $z < 6$ Ly$\alpha$ forest.  The cyan-solid and magenta-dashed faded curves show the ``fiducial'' and ``early'' models from~\citealt{Asthana2024}, and the dot-dashed gray curve shows the reference model\footnote{We actually show the reference model + ``extra un-resolved sinks'' (their Fig. 6) since it assumes the same sub-grid sinks opacity model we use in this work.  } from~\citealt{Cain2023}.  Our \textsc{late start/late end} model drops off faster than the others at $z > 9$, and our \textsc{early start/late end} model closely matches the ``early'' model from~\citealt{Asthana2024}.  The bottom panel of Figure~\ref{fig:emissivities} shows the integrated ionizing photon output of galaxies per H atom, with the vertical lines denoting the end of reionization and the horizontal lines the number of photons per H atom needed to complete reionization (the so-called ``ionizing photon budget'').  We find a budget of $\approx 2.6$, $2.8$, and $2.7$ photons per H atom for the \textsc{late start/late end}, \textsc{early start/late end}, and \textsc{early start/early end} models, respectively. 

All our models are mildly ``absorption-dominated'' - meaning $N_{\gamma/{\rm H}}$ is (slightly) more than twice the number of baryons per H atom in the universe ($1.082$, counting helium).  The photon budget depends mildly on the reionization history itself, but is determined mainly by the recombination rate predicted by our IGM opacity sub-grid model.  Recently,~\citealt{Davies2024b} made a first attempt to observationally measure the ``clumping factor'', which quantifies the recombination rate in the in-homogeneous IGM~\citep{Gnedin1997,Pawlik2009}.  They found $C \sim 12$ at $z > 5$, higher than the commonly assumed $C = 3$~\citep[e.g.][]{Finlator2011,McQuinn2011}.  For a reionization history similar to our \textsc{late start/late end} model, they find this higher $C$ increases the photon budget from $\approx 1.5$ to $\approx 3$.  Assuming that the number of recombinations, $N_{\gamma/{\rm H}} - 1.082$, is $\propto C$, the photon budget in the \textsc{late start/late end} model implies an effective $C \sim 9.5$, slightly below but consistent at $1\sigma$ with the~\citealt{Davies2024b} measurements\footnote{Caveats include (1) that the real effective clumping factor predicted by our model is likely not constant in redshift~\citep{DAloisio2020}, and (2) the integrated number of photons absorbed by the IGM is slightly smaller than the number produced, an effect not accounted for in~\citealt{Davies2024b}.  }.  Using this crude approximation, a recombination rate high enough to delay the end of the \textsc{early start/early end} model to $z \approx 5$ would require $C \sim 85$, an unrealistically high number even compared to the most extreme simulations\footnote{For $z_{\rm end} = 5.5$ ($6$), this estimation gives $C \sim 70$ ($55$).  }.  

\section{Numerical Methods} \label{sec:numerical}

In this section, we discuss some of the relevant technical details of our RT simulations, and our methods for forward-modeling the observables discussed in the next section.  The reader interested only in the results of this work may safely skip this section and pick up at \S\ref{sec:results}.  

\subsection{Radiative Transfer Simulations}
\label{subsec:RT}

We ran RT simulations of reionization using FlexRT (Cain \& D'Aloisio in prep.).  FlexRT is an adaptive ray tracing code that post-processes a time series of cosmological density fields to simulate the growth of ionized regions, the ionizing background, and the IGM thermal history.  The code uses a sub-grid model for opacity to ionizing photons in ionized gas, which captures the effects of $\sim$ kpc-scale structure that cannot be directly resolved.  Our sub-grid model is based on a suite of high-resolution, coupled hydro/RT simulations that can resolve the Jeans scale of cold, pre-ionized gas, run with the setup of~\citep{DAloisio2020,Nasir2021}.  We refer the reader to~\citep{Cain2021,Cain2022b}, and a forthcoming code paper (Cain \& D'Aloisio in prep.) for further details about FlexRT.  

Ionizing sources for the RT calculation are halos taken from a dark matter (DM)-only N-body simulation run with the particle-particle-particle-mesh (P$^3$M) code of~\citealt{Trac2015}.  This run has a box size of $L_{\rm box} = 200$ $h^{-1}$Mpc and $N = 3600^3$ DM particles, and uses a spherical over-density halo finder to identify halos.  We find that our halo mass function matches the halo mass function of~\citealt{Trac2015} to within $\approx 5-10\%$ (HMF) for $M_{\rm halo} > 3 \times  10^{9}$ $h^{-1}M_{\odot}$, and is $\approx 50\%$ incomplete at $M_{\rm halo} = 1 \times  10^{9}$ $h^{-1}M_{\odot}$.  To ensure we have a significant number of halos at the highest redshifts we simulate ($z > 15$), we adopt a mass cutoff for our sources of $M_{\rm halo} > 1 \times  10^{9}$ $h^{-1}M_{\odot}$, even though our HMF is incomplete there.  The density fields used in our RT calculation are taken from a high-resolution hydrodynamics simulation with the same large-scale initial conditions as the N-body run, which are re-binned to $N_{\rm RT} = 200^3$ for the RT.  Our simulations start at $z = 18$ and end at $z = 4.8$.  

We assign UV luminosities ($L_{\rm UV}$) to halos by abundance-matching to observed UV luminosity functions (UVLFs).  We use the measurements of~\citealt{Bouwens2021} at $z < 8$, those of~\citealt{Adams2024} at $8 < z < 14$, and that of~\citealt{Donnan2024} at $z = 14.5$ - those shown in the upper left panel of Figure~\ref{fig:UVLF_plot}.  At $z > 14.5$, we extrapolate the parameters of these UVLF fits to enable abundance matching up $z = 18$ - note that this choice has little effect on our results since none of our models are more than $2\%$ ionized at $z = 14.5$.  The ionizing emissivity is distributed between halos following $\dot{n}_{\gamma} \propto L_{\rm UV}$\footnote{The amounts to assuming that the product of the escape fraction and ionizing efficiency of galaxies, $f_{\rm esc}\xi_{\rm ion}$, is constant over the ionizing source population at a fixed redshift.   Note that although this is inconsistent with the model for $f_{\rm esc}$ and $\xi_{\rm ion}$ assumed in \S\ref{sec:JWST}, the spatial distribution of photons between sources is both highly uncertain and is a higher-order effect for the purposes of most of this work.  We will comment whenever it becomes relevant in subsequent sections.  }.  The volume-averaged ionizing emissivity of sources, $\dot{N}_{\gamma} = \frac{1}{L_{\rm box}^3} \sum_{i \in {\rm halos}} \dot{n}_{\gamma}^{i}$, is set by hand at each redshift and used to normalize the $\dot{n}_{\gamma}$ proportionality.  As explained in \S\ref{subsec:models}, we calibrated $\dot{N}_{\gamma}(z)$ for each scenario using a combination of JWST data (\S\ref{subsec:models}) and measurements from the $z < 6$ Ly$\alpha$ forest (\S\ref{subsec:lyaforest}, \S\ref{subsubsec:lya}).  We bin halos by the RT cell they occupy and treat cells containing one or more halos as sources.  We use a single ionizing frequency\footnote{See e.g.~\citealt[][]{Cain2023,Asthana2024} for discussions of the effect of multi-frequency RT on the Ly$\alpha$ forest.  } ($E_{\gamma} = 19$ eV), chosen to reproduce the same frequency-averaged HI cross-section, $\langle \sigma_{\rm HI} \rangle$, as a power law spectrum of the form $J_{\nu} \propto \nu^{-1.5}$ between 1 and 4 Rydbergs in the optically thin limit~\citep{Pawlik2011}.  Cells are assigned post ionization-front (I-front) temperatures using the flux-based method prescribed in~\citealt{DAloisio2019}.  The subsequent thermal history is calculated using their Eq. 6.  

\subsection{Modeling the Ly$\alpha$ forest}
\label{subsec:lyaforest}

We model the Ly$\alpha$ forest in post-processing using the density fields from our aforementioned high-resolution hydrodynamics simulation, which has $N = 2048^3$ gas cells.  Ionized fractions, photo-ionization rates ($\Gamma_{\rm HI}$) and temperatures are mapped from the FlexRT simulations onto these density fields, and the residual neutral fraction in ionized regions is calculated assuming photo-ionization equilibrium and the case A recombination rate\footnote{Case A is appropriate for the under-dense gas that set the forest transmission at these redshifts.  }.  The native spatial resolution of our density field is $\Delta x_{\rm cell} = 97.6$ $h^{-1}$kpc, too coarse to fully resolve the low-density voids that set the transmission at $5 < z < 6$~\citep{Doughty2023}.  We apply the multiplicative correction prescribed in Appendix A of~\citealt{DAloisio2018} to our residual neutral fractions to get an effective resolution of $\Delta x_{\rm cell} = 12.2$ $h^{-1}$kpc.  The Ly$\alpha$ voigt profile is approximated using the analytic fit from~\citealt{TepperGarcia2006}.  

Since our gas temperatures are calculated on the coarse RT grid, we do not capture the temperature-density relation on scales smaller than the RT cell size.  As temperature (usually) correlates positively with density in the IGM, low (high)-density hydro cells embedded in larger RT cells will be assigned temperatures that are too high (low).  To correct for this, we assign a local temperature-density relation to each cell using the procedure described in~\citealt{Cain2023} (see also their Appendix E), which uses the IGM temperature model of~\citealt{McQuinn2016}.  We find this lowers the mean transmission by $\approx 10-15\%$, since the correction cools the under-dense cells, which affect the mean transmission the most.  

We compute Ly$\alpha$ forest statistics by casting $4000$ sightlines from random locations and in random directions of length $50$ $h^{-1}$Mpc, for a total path length of $200$ $h^{-1}$Gpc.  We calculate transmission statistics from $z = 4.8$ to $z = 6$ in $\Delta z = 0.2$ increments.  Since our native resolution of $97.6$ $h^{-1}$kpc is only $2-3\times$ narrower than the typical width of Ly$\alpha$ line profile ($\approx 15$ km/s vs. $30-50$ km/s), we do the integration over the line at a velocity resolution $4 \times$ higher than that of the hydro simulation.  We find this reduces the mean transmission by a few percent at most.   

\subsection{Ly$\alpha$ transmission around galaxies}
\label{subsec:lyatransmission}

We have also modeled Ly$\alpha$ transmission on the red side of line center around halos that could host Ly$\alpha$ emitting galaxies (LAEs). 
This allows us to assess the statistics of LAE visibility in our models at $z > 7$.  LAEs typically emit Ly$\alpha$ red-shifted to both the red and blue sides of line center~\citep{Verhamme2006}.  Although any emission on the blue side would be absorbed by even the ionized part of the IGM at these redshifts, attenuating the red side requires damping wing absorption from the fully neutral IGM~\citep[see][and references therein]{Mason2020}.  This makes LAEs a potentially powerful probe of the IGM neutral fraction.  However, interpreting observed red-side Ly$\alpha$ emission (or lack thereof) is complicated by uncertainties in modeling the intrinsic line profile of the LAEs themselves, and other factors such as surrounding inflows/outflows and proximate self-shielding systems~\citep{Park2021,Smith2022}.   

In this work, we will avoid modeling the intrinsic LAE line profile, instead focusing on the IGM transmission, $T_{\rm IGM}$.  To calculate this, we trace $50$ randomly oriented sightlines away from each halo and compute the Ly$\alpha$ transmission profiles along each sightline at $\pm 5$ $\text{\AA}$ from line center ($\approx \pm 800$ km/s).  We use the same $N = 2048^3$ high-resolution hydro simulation for this calculation as for the Ly$\alpha$ forest, and we calculate the ionization state of the ionized gas in the same way\footnote{Our results are largely insensitive to how the highly ionized IGM is modeled, however, since the red side transmission is set by the damping wing from the neutral IGM.  }.  We begin integrating the Ly$\alpha$ opacity $500$ $h^{-1}$kpc (5 hydro cells) away from the location of the halo to avoid gas within the halo itself contributing to $T_{\rm IGM}$.  Gas velocities relative to the halo are computed by subtracting the halo velocity measured from the N-body simulation.  

The gas around massive halos can have sharp line-of-sight velocity gradients owing to inflows near the halo.  Sightlines pointing away from these halos see positive velocity gradients, which narrows the Ly$\alpha$ line in redshift space.  We find that the resulting sharp jumps in velocity between adjacent cells can produce artifacts in our transmission spectra.  To mitigate this, we linearly interpolate the gas velocities (and all other relevant quantities) onto a grid with $4 \times$ higher resolution than that of the simulation when calculating $T_{\rm IGM}$ (see~\citealt{Gangolli2024} for a description of a similar procedure).  We find the transmission profiles to be well-converged at this resolution.  

\subsection{Modeling the Patchy kSZ signal}
\label{subsec:model_pksz}

CMB photons can scatter off free electrons during reionization.  If the electrons are moving relative the the CMB rest frame, this results in a Doppler shift of the photons.  This can shift the blackbody spectrum and result in additional temperature anisotropies in the CMB \citep{Sunyaev1980-pq}. This is known as the Sunyaev-Zel'dovich Effect, and its resultant temperature deviation along a line of sight is given by:

\begin{equation}
    \label{eq:dtq}
    \frac{\Delta T}{T} = - \sigma_T \overline{n}_{e,0} \int e^{-\tau_{\rm es}} \frac{\hat{\gamma}\cdot\vec{q}}{c} \frac{ds}{a^2} 
\end{equation}

where $\hat{\gamma}$ is the line of sight direction, $\vec{q}$ = (1 + $\delta) \chi \vec{v}$ is the ionized momentum field, $\sigma_T$ is the Thompson scattering cross section and $\tau_{\rm es}$ is the CMB optical depth to the last scattering surface, and $\overline{n}_{e,0}$  is the mean electron density at $z$=0.  The integral can be broken up into a post-reionization, homogeneous kinetic Sunyaev-Zel'dovich Effect (hkSZ), and a high-$z$ patchy kinetic Sunyaev-Zel'dovich Effect (pkSZ) while reionization is still occurring.  Untangling these components is difficult, so observations use templates to subtract the hkSZ component from the measured total kSZ power.  For the purposes of this work, we take any contributions from $z \gtrsim$ 5 as part of the pkSZ, even in simulations where reionization ends at $z >$ 5.  This keeps the different scenarios directly comparable to each other.  Note that actual measurements (such as the one from SPT by~\citealt{Reichardt2020}, see \S\ref{subsec:cmb}) must assume a fixed $z_{\rm end}$ in their analysis.  

We calculate the signal using a method similar to that first suggested by~\citealt{Park2013}.  The kSZ angular power spectrum can be calculated from the 3D power spectrum of $\vec{q}$ by: 

\begin{equation}
    \label{eq:Clpq}
    C_{\ell} = \left( \frac{\sigma_T \overline{n}_{e,0}}{c} \right)^2 \int \frac{ds}{s^2 a^4} e^{-2\tau} \frac{P_{q_{\perp}} \left(k = \ell/s, s\right)}{2}   
\end{equation}
where $ (2\pi )^3 P_{q_{\perp}}(k) \delta^{D}(\vec{k}-\vec{k'})=   \langle{\widetilde{q}_{\perp}(\vec{k}) 
  \cdot \widetilde{q}_{\perp}^* (\vec{k'})}  \rangle  $ 
is the power spectrum of the specific ionized momentum modes perpendicular to the Fourier wave vector, $\widetilde{q}_{\perp} = \widetilde{q} - \hat{k} (\widetilde{q} \cdot \hat{k} )$, $\delta^{D}(k-k')$ is the Dirac-$\delta$ function and
$\widetilde{q} = \int \vec{q} e^{-i \vec{k} \cdot \vec{q}} d^3 \vec{x} $ denotes the Fourier transform of $\vec{q}$.  Only the $\widetilde{q}_{\perp}$ mode contributes significantly to the kSZ signal, due to $\widetilde{q}_{||}$ contributions canceling out when integrating over the line of sight.  We refer the reader to \citealt{Ma2002} and appendix A in \citealt{Park2013} for details.  In general, it is common to write the angular power spectrum in the dimensionless form: $D_{\ell} = C_{\ell} ( \ell + 1 ) \ell / (2 \pi)$.

Running RT simulations imposes practical limitations on the volume of our box. This limited box size means we miss large scale velocity modes that add to the kSZ power.  To compensate for this, we use the same correction applied in \citealt{Park2013} eq. (B1), and take advantage of linear theory to generate a correction term:

\begin{equation}
    \label{eq:PQmiss}
     P_{q_{\perp}}^{{\rm miss}}(k,z) = {\int_{k<k_{{\rm box}}}} \frac{d^3 k'}{(2\pi)^3} (1-\mu^2) P_{\chi(1+\delta)}( | \vec{k} - \vec{k'} | ) P_{vv}^{{\rm lin}} ( k')     
\end{equation}
where $k_{\text{box}} = 2 \pi / L_{\text{box}}$ is the box-scale wavemode, $\mu = \hat{k} \cdot \hat{k'}$, $P_{\chi(1+\delta),\chi(1+\delta)}(k)$ is the ionized matter power spectrum and $P_{vv}^{\text{lin}}(k) = (\dot{a}f/k) P_{\delta \delta}^{\text{lin}}(k)$ is the velocity power spectrum in the linear approximation taken from the public code CAMB \citep{Lewis2000}.  

It was found by \cite{Alvarez2016} that this kind of approach can still underestimate the $D^{\text{pkSZ}}_{3000}$ by $\sim$10-20$\%$ due to the irreducible or connected component of the $\langle \delta_{\chi} v \cdot \delta_{\chi}v\rangle$ term being non-negligible because of the non-Gaussianity of reionization.  So our calculations could be seen as conservative estimates of the power.  As we will see in \S\ref{subsec:cmb}, an increase in power could strengthen our conclusions.  As such, we do not expect this missing term to affect our qualitative results.  

\vspace{0.5cm}

\section{Implications of Other Reionization Observables} \label{sec:results}

In the rest of this work, we study three other observational windows into reionization: measurements from the spectra of high-redshift QSOs at $z \leq 6.5$ (\S\ref{subsec:QSO}), observations of Ly$\alpha$-emitting galaxies at $z \geq 8$ (\S\ref{subsec:LAEs}), and the patchy kSZ effect from reionization (\S\ref{subsec:cmb}).  Our goal will be to see if these observations, together with aforementioned JWST data, reveal a consistent picture about when reionization started and when it ended.  

\subsection{QSO Observations at $5 < z < 6$} \label{subsec:QSO}

\subsubsection{The Ly$\alpha$ Forest} \label{subsubsec:lya}

The $5 < z < 6$ Ly$\alpha$ forest is perhaps the most compelling indicator that reionization ended at $z < 6$.  This conclusion has emerged from studies of the mean transmission of the Ly$\alpha$ forest and the scatter in Ly$\alpha$ opacities~\citep{Kulkarni2019,Keating2019,Nasir2020,Bosman2021}.  We will study both in this section.  

\begin{figure}
    \centering
    \includegraphics[scale=0.22]{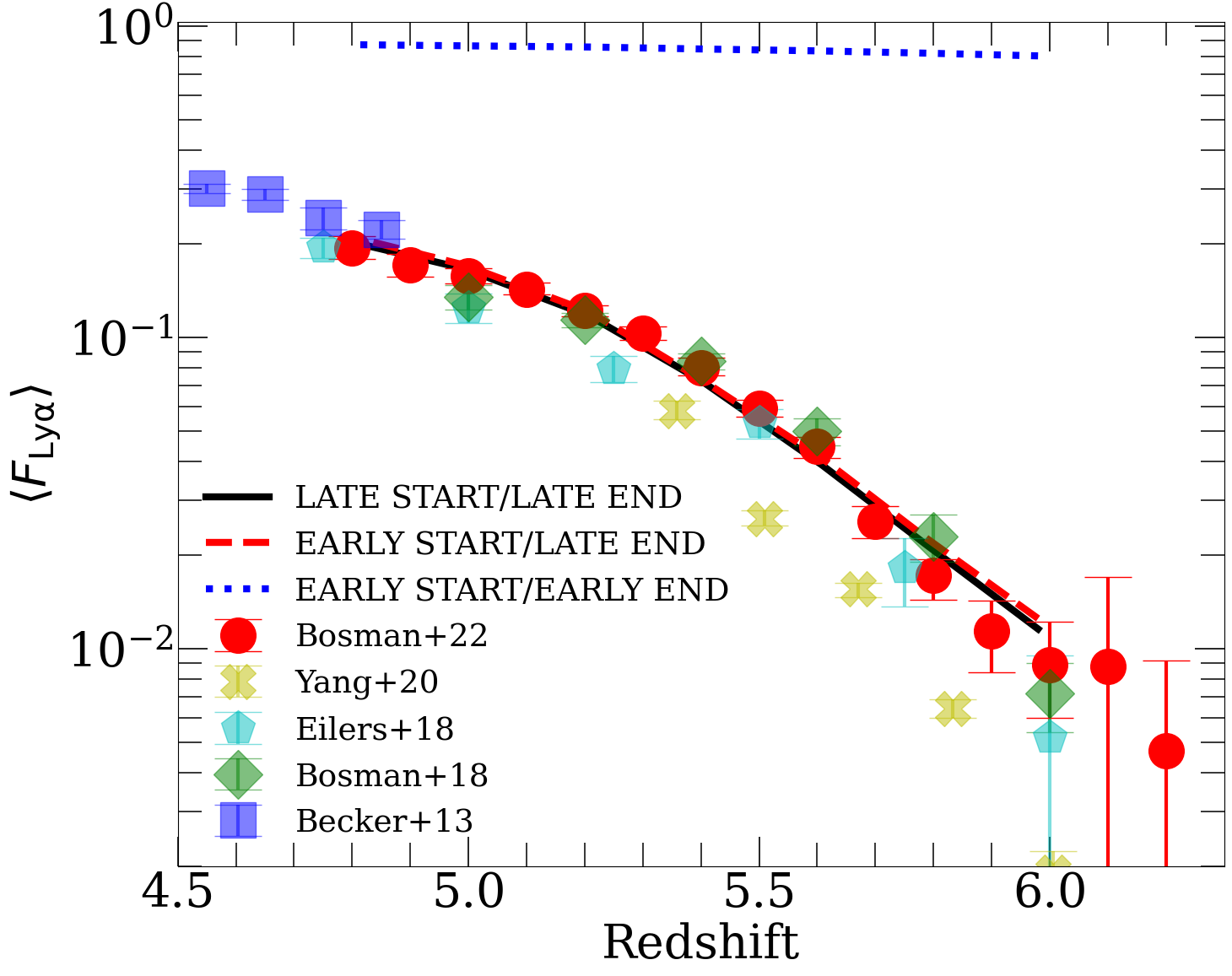}
    \caption{Ly$\alpha$ forest mean transmission, compared with recent observations (see text).  The \textsc{early start/early end} model is in severe disagreement with measurements, predicting transmission near unity at all redshifts.  The other two models agree with the measurements of~\citealt{Bosman2021} by construction, since their $\dot{N}_{\gamma}(z)$ histories were calibrated to match those measurements (see discussion in \S\ref{subsec:models}).  The fact that such a calibration is possible for both late-ending models indicates that $\langle F_{\rm Ly\alpha} \rangle$ alone cannot distinguish between them. 
 }
    \label{fig:qso_summary_1}
\end{figure}

Figure~\ref{fig:qso_summary_1} shows the mean transmission of the Ly$\alpha$ forest, $\langle F_{\rm Ly\alpha} \rangle$, at $5 \leq z \leq 6$ in our models compared with recent measurements from~\citealt{Becker2013,Eilers2018,Bosman2018,Yang2020b,Bosman2021}.  The \textsc{late start/late end} and \textsc{early start/late end} models both agree well with the data - this is by design, since both $\dot{N}_{\gamma}$ histories were calibrated so that the simulations would match the~\citealt{Bosman2021} measurements of $\langle F_{\rm Ly\alpha} \rangle$.  As such, these two models cannot be distinguished using the mean forest transmission alone.  We see that the \textsc{early start/early end} model misses the measurements severely, predicting a mean transmission near unity at all redshifts.  Indeed, the average HI photo-ionization rate, $\Gamma_{\rm HI}$, in ionized gas is $\approx 2.7 \times 10^{-11}$ ${\rm s}^{-1}$ at $z = 6$, $\approx 2$ orders of magnitude higher than measurements at that redshift~\citep{DAloisio2018,Gaikwad2023}.  Thus, the forest transmission measurements strongly disfavor a $z \sim 8$ end to reionization.  Because of this, we will omit the \textsc{early start/early end} model from many of our subsequent comparisons, and focus instead on whether observations can distinguish the other two scenarios.  

\begin{figure*}
    \centering
    \includegraphics[scale=0.19]{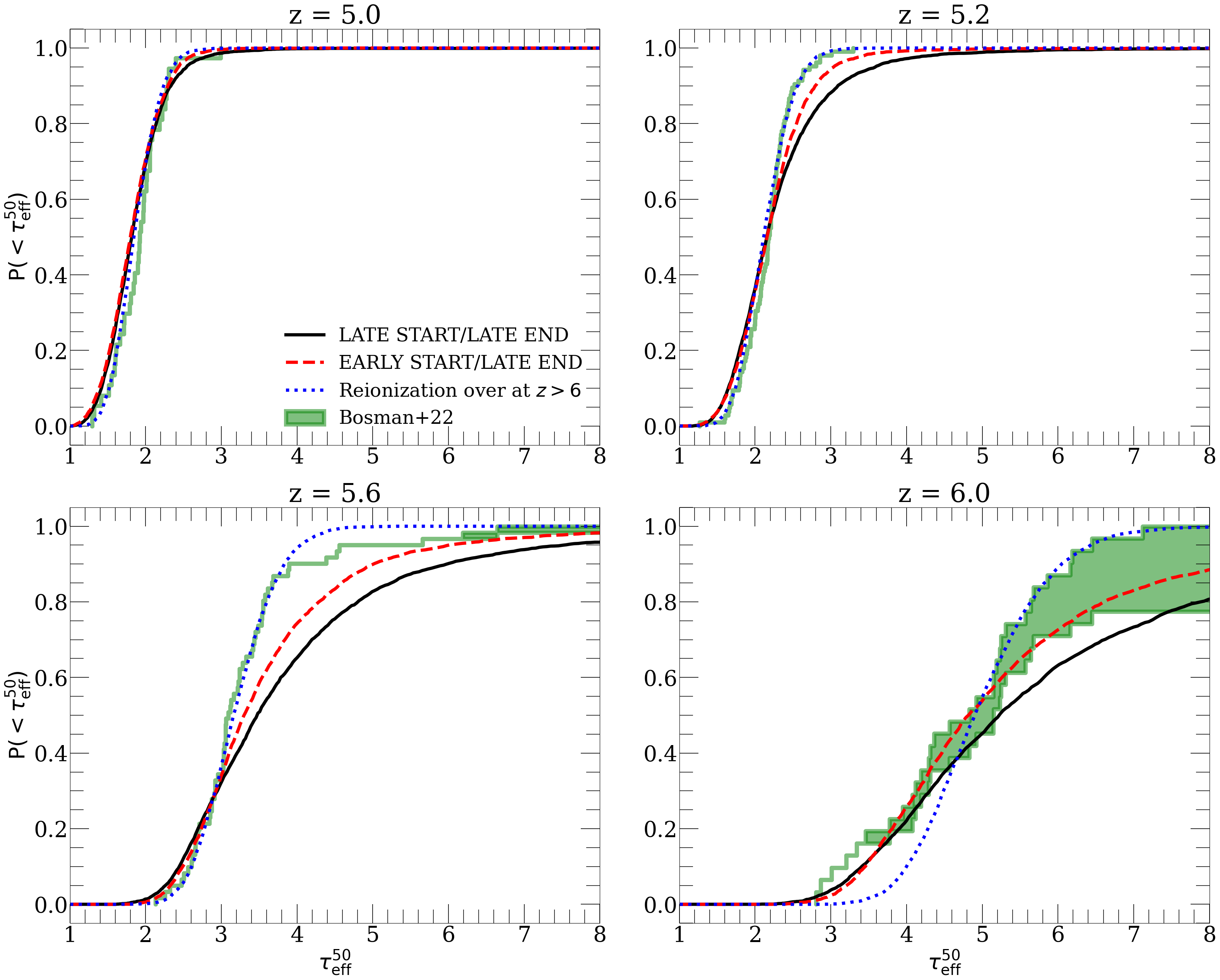}
    \caption{Distribution of effective Ly$\alpha$ optical depths over $50$ $h^{-1}$Mpc segments of the Ly$\alpha$ forest, $P(< \tau_{\rm eff}^{50})$, compared to measurements from~\citealt{Bosman2021}.  We show results and measurements at $z = 5$, $5.2$, $5.6$, and $6$ (see panel titles).  For the \textsc{early start/early end} model, we have re-scaled the Ly$\alpha$ opacities along each all sightlines by a constant value such that the mean transmission matches the~\citealt{Bosman2021} measurements. 
    This allows us to make a clean comparison of how the IGM neutral fraction affects the shape of $P(< \tau_{\rm eff}^{50})$ at fixed $\langle F_{\rm Ly\alpha} \rangle$.   At $z = 6$, the re-scaled \textsc{early start/early end} model produces too little scatter in $\tau_{\rm eff}^{50}$, the \textsc{late start/late end} case produces too much, and the \textsc{early start/late end} model is a good match.  At $z = 5.6$, none of the models match the observations particularly well.  The observations lie between the blue dotted and red dashed curves, suggesting a non-zero neutral fraction smaller than that of the \textsc{early start/late end} model ($7.5\%$).  At $z = 5.2$, the re-scaled \textsc{early start/early end} matches the data best, suggesting that reionization should end slightly earlier than it does in both late-ending models.  All models match the data at $z = 5$, when reionization is complete.  }
    \label{fig:taueff}
\end{figure*}

Figure~\ref{fig:taueff} shows the cumulative distribution function (CDF) of forest effective optical depths measured over intervals of $50$ $h^{-1}$Mpc, $P(< \tau_{\rm eff}^{50})$.  The $\tau_{\rm eff}^{50}$ distribution contains more information than $\langle F_{\rm Ly\alpha} \rangle$, since it is sensitive to the spatial fluctuations in the IGM ionization state.  This shape is sensitive to the IGM neutral fraction, since neutral islands produce high-$\tau_{\rm eff}^{50}$ sightlines at $z < 6$ in late reionization scenarios~\citep{Kulkarni2019,Nasir2020,Qin2021}.  We show $P(< \tau_{\rm eff}^{50})$ at $z = 5$, $5.2$, $5.6$, and $6$ for each of our models, alongside measurements from~\citealt{Bosman2021}.  Note that at $z = 5.6$ and $6$, the shaded green regions denote the spread between optimistic and pessimistic assumptions about non-detections of transmitted flux in opaque sightlines - see~\citet{Bosman2021} for details.  To quantitatively compare our models to the data, we evaluate the Cramer-von Mises distance statistic, $\delta P(<\tau_{\rm eff}^{50}) = N\int_{0}^{1} dP_{\rm sim}(<\tau_{\rm eff}^{50}) \left[P_{\rm sim}(<\tau_{\rm eff}^{50}) - P_{\rm data}(<\tau_{\rm eff}^{50})\right]^2$, where $N$ is the number of measurements in the data sample.  For $P_{\rm data}(<\tau_{\rm eff}^{50})$, we use the average of the optimistic and pessimistic scenarios reported by~\citet{Bosman2021}. 
 We tabulate $\delta P(<\tau_{\rm eff}^{50})$ for all of our models and redshifts in Table~\ref{tab:deltaPtau} in Appendix~\ref{app:Ptau}, and give some key values in the discussion below.    

It is instructive to compare our late-ending models to a hypothetical scenario in which reionization is completely over by $z = 6$, but for which $\langle F_{\rm Ly\alpha} \rangle$ agrees with measurements.  Such a comparison allows us to cleanly assess the effects of spatial fluctuations in $\Gamma_{\rm HI}$ and neutral islands on $P(<\tau_{\rm eff}^{50})$.  We generate such a scenario by re-scaling $\Gamma_{\rm HI}$ in the \textsc{early start/early end} model until $\langle F_{\rm Ly\alpha} \rangle$ agrees with the measurements.  We emphasize that we do not consider this a realistic, physical scenario\footnote{Indeed, the re-scaling factors used here were $\sim 10^2$ at all redshifts!  }, but rather as a means to aid interpretation of the results for the two late-ending models.  This is shown by the blue dotted curves in Figure~\ref{fig:taueff}.  

At $z = 6$ (lower right),  the \textsc{early start/late end} model best matches the~\citealt{Bosman2021} data.  The blue dotted curve shows that a $z > 6$ end to reionization results in a CDF that is too narrow, implying too little scatter in $\tau_{\rm eff}^{50}$ ($\delta P(<\tau_{\rm eff}^{50}) = 0.275$), echoing the finding of~\citealt{Bosman2021} that a homogeneous UVB cannot explain the observed scatter.  Conversely, the \textsc{late start/late end} model predicts {\it too much} scatter (the CDF is too wide, with $\delta P(<\tau_{\rm eff}^{50}) = 0.179$).  This is because it has a neutral fraction of $\approx 30\%$ at $z = 6$, as compared to only $15\%$ in the \textsc{early start/late end} case.  The \textsc{early start/late end} model shows the best agreement, with $\delta P(<\tau_{\rm eff}^{50}) = 0.0438$.  At $z = 5.6$ (lower left), neither of hte late-ending models match the observations very well.  Both produce too much scatter in $\tau_{\rm eff}^{50}$, with $\delta P(<\tau_{\rm eff}^{50}) = 1.314$ and $\delta P(<\tau_{\rm eff}^{50}) = 0.589$ for the late and early start, respectively. By contrast, the early-ending scenario has too little scatter at the high-$\tau$ end, but is a much better fit to the data overall, with $\delta P(<\tau_{\rm eff}^{50}) = 0.0565$.  This suggests that a model with a non-zero neutral fraction smaller than that in the \textsc{early start/late end} case ($7.5\%$) would match the data best.  At $z = 5.2$ (upper right), the early-ending scenario fits the data better than the other two ($\delta P(<\tau_{\rm eff}^{50}) = 0.339$), which both have too much scatter - $\delta P(<\tau_{\rm eff}^{50}) = 0.865$ and $\delta P(<\tau_{\rm eff}^{50}) = 0.4746$ for the late and early start, respectively.  This indicates that both late-ending models may end reionization slightly too late, but that the \textsc{early start/late end} model consistently fits the observations better.  At $z = 5$ (upper left), when the neutral fraction is $< 1\%$ in all three models, they agree well with the observations.  

At face value, observations of $P(< \tau_{\rm eff}^{50})$ at $5 \leq z \leq 6$ seem to prefer the \textsc{early start/late end} model.  This scenario has a lower neutral fraction at $z \leq 6$ than the \textsc{late start/late end} case, and as such better matches the observed scatter in $P(< \tau_{\rm eff}^{50})$.  However, {\it none} of the scenarios in Figure~\ref{fig:taueff} match the observed $P(< \tau_{\rm eff}^{50})$ at all redshifts.  This is likely because our late-ending models finish reionization too late, as evidenced by the $z = 5.2$ comparison (upper right panel).  The findings of~\citealt{Bosman2021}, based on these same measurements, suggest that reionization should be complete by $z = 5.3$ (vs. $z \approx 5$ in our models), a shift of $\Delta z \approx 0.3$.  In Appendix~\ref{app:Ptau}, we estimate what $P(< \tau_{\rm eff}^{50})$ would look like if both late-ending models finished reionization at $z = 5.3$.  We use the FlexRT outputs at $z' = z - 0.3$ to calculate $P(< \tau_{\rm eff}^{50})$, then re-scale $\Gamma_{\rm HI}$ in ionized gas until $\langle F_{\rm Ly\alpha} \rangle$ matches measurements.  We show that this procedure brings our simulations into better agreement with the measurements, and that the \textbf{\textsc{early start/late end}} model remains preferred.  This finding suggests that our conclusion about which scenario is preferred by $P(< \tau_{\rm eff}^{50})$ is reasonably robust to the possibility that our reionization ends slightly too late in our models.  

In~\citealt{Cain2023}, we pointed out several factors that can affect the precise timing of reionization's end in models matched to measurements of $\langle F_{\rm Ly\alpha} \rangle$, and the strength of the associated $\tau_{\rm eff}^{50}$ fluctuations.  First, lack of spatial resolution in the forest can lead to an under-estimate of the mean transmission at fixed $x_{\rm HI}$~\citep{Doughty2023}, resulting in a spuriously early end to reionization (by $\Delta z \approx 0.2$, see Fig. 11 of~\citealt{Cain2023}) when calibrating to measurements.  Our forest calculations include the resolution correction prescribed Appendix A of~\citealt{DAloisio2018}, so in principle they account for this effect.  However, those corrections were derived in a $25$ $h^{-1}$Mpc box, and it is unclear how they might change in a larger box.  It is therefore possible that we have over-corrected for resolution.  We also found in~\citealt{Cain2023} that harder ionizing spectra result in an earlier end to reionization at fixed $\langle F_{\rm Ly\alpha} \rangle$.  

Another major uncertainty is the clustering of ionizing sources, which affects large-scale fluctuations in the ionizing background and the structure of neutral islands.  Models driven by rare, highly clustered sources (such as AGN~\citep{Chardin2015,Madau2024}), have stronger fluctuations in $\Gamma_{\rm HI}$ near the end of reionization, leading to a wider $P(\tau_{\rm eff})$. However, less clustered sources lead to more extended, porous neutral islands, causing stronger $\tau_{\rm eff}$ fluctuations when reionization is ongoing.  We further showed in~\citet{Cain2023} that, at fixed $\langle F_{\rm Ly\alpha}\rangle$, models driven by less clustered sources end reionization earlier, which would bring our $P(<\tau_{\rm eff}^{50})$ results into better agreement with observations (see also~\citet{Asthana2024}).  In light of these caveats, we emphasize that our results should not be taken to show that reionization ended at exactly $z = 5$ - just that it must have ended below $z = 6$.  Exploring these effects in detail would require a much larger number of simulations, and is beyond the scope of this work.  

\subsubsection{The Mean Free Path} \label{subsubsec:mfp}

\begin{figure*}
    \centering
    \includegraphics[scale=0.25]{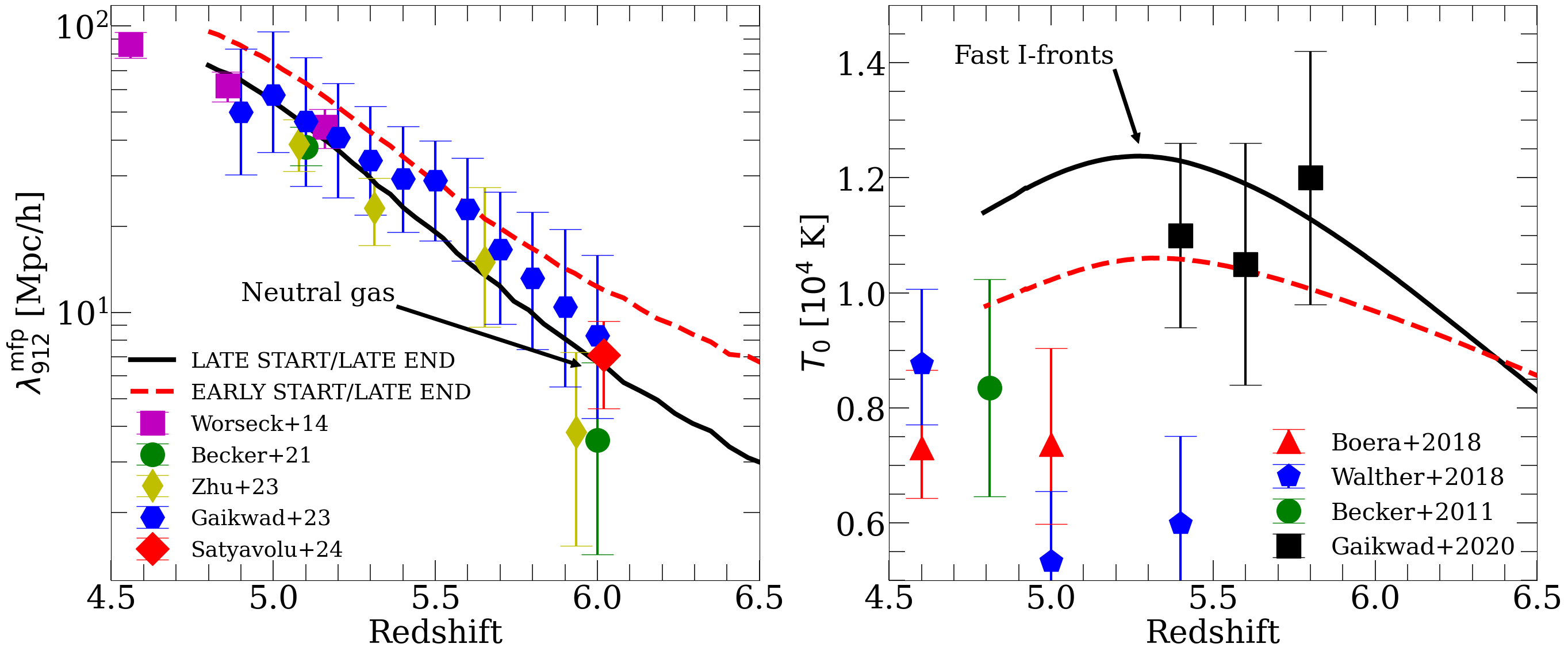}
    \caption{Other (non-Ly$\alpha$ forest) QSO-based measurements of global IGM properties at $5 < z < 6$ in our late-ending models.  {\bf Left:} the mean free path to ionizing photons at $912$ $\text{\AA}$ compared to recent measurements.  We show direct measurements from the Lyman Continuum spectra of QSOs~\citep{Worseck2014,Becker2021,Zhu2023,Satyavolu2023}, and the recent indirect measurements of~\citealt{Gaikwad2023} based on the Ly$\alpha$ forest.  Both models are consistent with the forest-based measurements from~\citealt{Gaikwad2023}, but the direct QSO-based measurements (particularly at $z = 6$) prefer the \textsc{late start/late end} scenario.  The faster redshift evolution of $\lambda_{\rm mfp}$ in that model, driven by the evolving neutral fraction, is in better agreement with the $z = 6$ direct measurements.  {\bf Right:} Average temperature at mean density ($T_0$), compared to measurements from~\citealt{Becker2011,Boera2019,Walther2019,Gaikwad2020}.  Both models display reionization-driven temperature peaks at $z \sim 5.3$, consistent with the~\citealt{Gaikwad2020} measurements.  The \textsc{late start/late end} model peaks at a higher temperature, since a larger fraction of the IGM has been recently heated by fast-moving I-fronts (see annotation).  The \textsc{early start/late end} model is more consistent with measurements at $z \geq 5$, indicating a mild preference for that scenario (but with some caveats - see text).  }
    \label{fig:qso_summary_2}
\end{figure*}

Next, we will study the mean free path to ionizing photons (MFP, $\lambda_{\rm mfp}$), the average distance an ionizing photon travels through the IGM before being absorbed.  The MFP is sensitive to the distribution of neutral gas in the IGM and small-scale clumping in the ionized IGM~\citep{Emberson2013,Park2016,DAloisio2020,Chan2023}.   We calculate the Lyman-limit MFP in our simulations using the definition in Appendix C of~\citealt{Chardin2015}, 
\begin{equation}
    \lambda_{\rm mfp}^{912} = \frac{\langle \int x df \rangle}{\langle \int df \rangle} = - \left\langle\int_{1}^{0} x df \right\rangle
\end{equation}
where $x$ is the position along a randomly oriented sightline, $f(x)$ is the transmission of $912\text{\AA}$ photons, and the angle brackets denote an average over many sightlines. \citealt{Roth2023} found that this definition matches well with forward-modeled direct MFP measurements from QSO spectra, even in a partially neutral IGM.  We caution, however, that different ways of estimating the MFP from simulations can give modestly different results, especially when reionization is ongoing~\citep{Lewis2022,Satyavolu2023}.  

The left panel of Figure~\ref{fig:qso_summary_2} shows the MFP in our late-ending models, compared to measurements from~\citealt{Worseck2014,Becker2021,Zhu2023,Gaikwad2023}.  Both scenarios are in broad agreement with the measurements.  However, the direct measurements using QSO Lyman Continuum (LyC) spectra (all but the~\citealt{Gaikwad2023} points) display a preference for the \textsc{late start/late end} model.  This is largely due to the short $z = 6$ direct measurements, which prefer the rapid neutral fraction-driven decline in $\lambda_{\rm mfp}^{912}$ in the \textsc{late start/late end} case.  By contrast, the \textsc{early start/late end} model is $\approx 2\sigma$ away from the central values of the $z = 6$ measurements from~\citealt{Becker2021,Zhu2023}.  Both models are within $1\sigma$ of the indirect, Ly$\alpha$ forest-based measurements from~\citealt{Gaikwad2023} (see also~\citet{Davies2024}). 

This result is consistent with previous findings that ongoing reionization at $z = 6$ is needed to explain the direct QSO measurements~\citep{Cain2021,Lewis2022,Garaldi2022,Lewis2022,Satyavolu2023,Roth2023}.  Another effect at play is that the ionized IGM at $z = 6$ in the \textsc{late start/late end} was more recently ionized, and thus clumpier~\citep{Park2016,DAloisio2020}.  These factors result in direct MFP measurements preferring the \textbf{\textsc{late start/late end}} scenario.  Our earlier result that $P(< \tau_{\rm eff}^{50})$ measurements prefer the \textsc{early start/late end} model hints at a possible tension between the Ly$\alpha$ forest and direct MFP measurements.  This is consistent with the fact that the indirect MFP measurements from~\citealt{Gaikwad2023} (based on $P(< \tau_{\rm eff}^{50})$ itself) at $z = 6$ are a factor of $\approx 2$ above the direct measurements.  We emphasize that this tension is mild, since the $1 \sigma$ error bars of these measurements overlap.   

\subsubsection{IGM Thermal History} \label{subsubsec:thermal}

The IGM temperature at mean density, $T_0$, is shown in the right panel of Figure~\ref{fig:qso_summary_2}, alongside measurements from~\citealt{Becker2011,Boera2019,Walther2019,Gaikwad2020}.  Both late-ending models display a ``bump'' in temperature at $z \approx 5.3$ due to the end of reionization.  Heating by I-fronts~\citep{DAloisio2019,Zeng2021} increases $T_0$ until near reionization's end, after which cooling from the expansion of the universe and Compton scattering off the CMB set the evolution of $T_0$~\citep{McQuinn2016}.  The peak in $T_0$ is higher in the \textsc{late start/late end} model for two reasons.  First, the IGM has reionized more recently on average, and thus has had less time to cool.  Second, ionization fronts in that model move faster on average since reionization takes place over a shorter period of time, resulting in higher post I-front temperatures~\citep{DAloisio2019}.  
The redshift of the bump suggested by the~\citealt{Gaikwad2020} measurements is closer to $z \sim 5.6$ - this is consistent with our earlier finding (based on $P(< \tau_{\rm eff}^{50})$) that reionization may end $\Delta z \sim 0.3$ too late in our models.  

At face value, the \textsc{early start/late end} model agrees best with $T_0$ measurements.  Although both models are consistent with the~\citealt{Gaikwad2020} points, the \textsc{late start/late end} is too hot at $z \geq 5$ for the measurements there.  In the \textsc{early start/late end} case, a larger fraction of the IGM has re-ionized at higher redshift, giving it more time to cool by $z = 5$.  That model also agrees well with the reionization history in the best-fitting model of~\citealt{Villasenor2022}, which fits a broad range of IGM temperature measurements down to $z = 2$.  That model has ionized fractions of $\approx 35\%$ ($\approx 15\%$) at $z = 8$ ($10$), similar to our \textsc{early start/late end} scenario (which has $1 - x_{\rm HI} \approx 40\%$, ($20\%$)).  An important caveat is that the thermal history is sensitive at the $20-30\%$ level to the spectrum of the ionizing radiation, through a combination of the post I-front temperature ($T_{\rm reion}$) and photo-heating in ionized gas afterwards~\citep{DAloisio2019}.  For example, a much softer ionizing spectrum could shift the $T_0$ histories significantly lower at fixed reionization history (see e.g. the bottom middle panel of Fig. 3 in~\citealt{Asthana2024}).  This could bring the \textsc{late start/late end} model into agreement with $z \leq 5$ $T_0$ measurements.  However, this would also require a later reionization history at fixed $\langle F_{\rm Ly\alpha} \rangle$~\citep[Figure 5 of][]{Cain2023}, which would worsen the disagreement with the observed $P(< \tau_{\rm eff}^{50})$ in Figure~\ref{fig:taueff}.  As such, we conclude that measurements of $T_0$ mildly prefer the \textbf{\textsc{early start/late end}} model.  

\subsubsection{Neutral fraction constraints at $z \leq 6.5$}
\label{subsubsec:QSOxHI}

\begin{figure}
    \centering
    \includegraphics[scale=0.19]{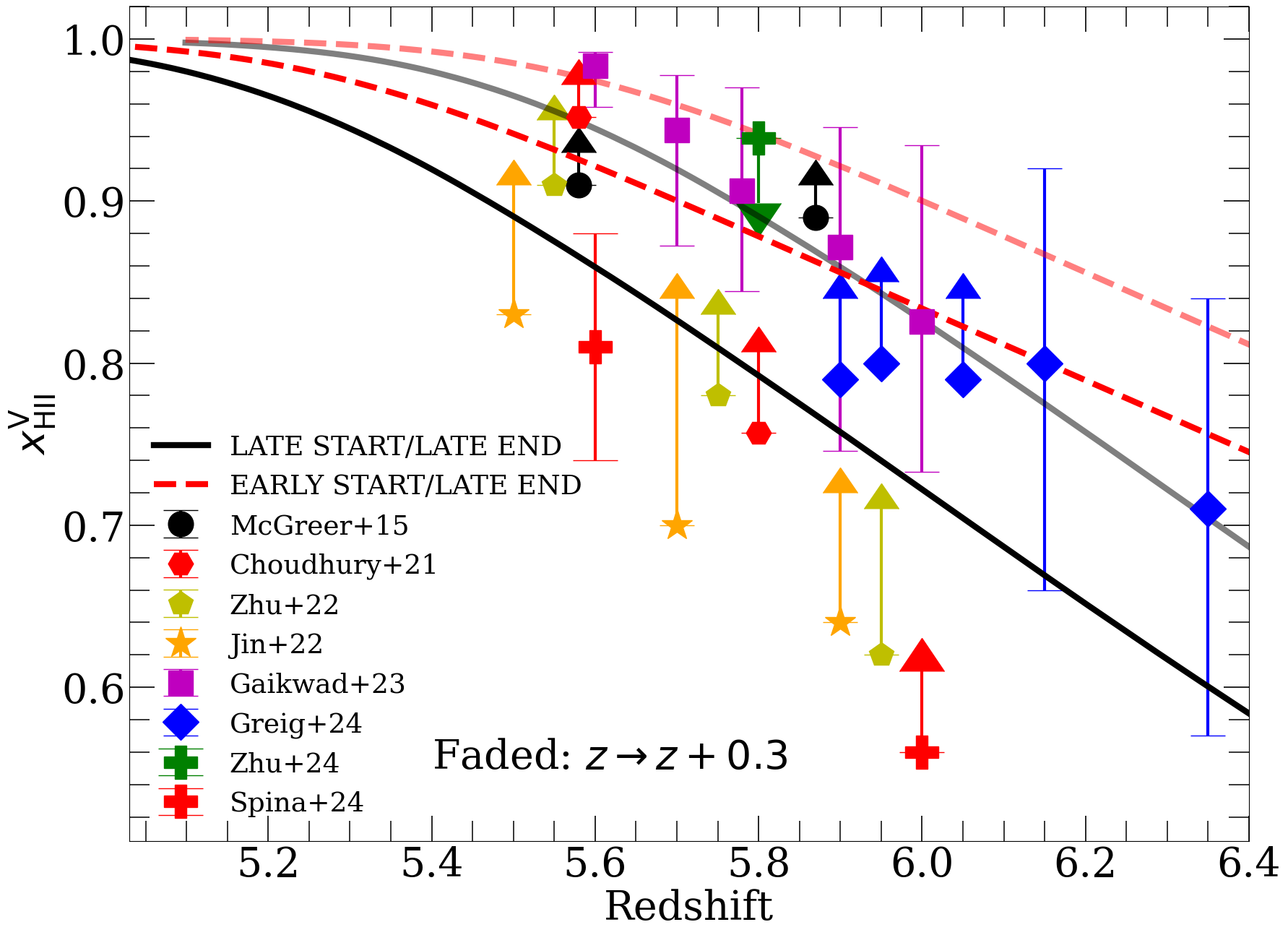}
    \caption{Constraints on the neutral fraction at $z \leq 6.5$ from dark pixels, dark gaps, damping wings, $P(< \tau_{\rm eff}^{50})$, and forest damping wings (see text for references).  The bold curves show our late-ending reionization histories, and the faded curves show these shifted to the right by $\Delta z = 0.3$ (see \S\ref{subsubsec:lya}).  The upper limit at $z = 5.8$ from~\citealt{Zhu2024} from Ly$\alpha$ forest damping wings is inconsistent with an earlier ending to reionization, and the same can be said of the $z = 5.6$ measurement from~\citealt{Spina2024}.  Some of the lower limits on the ionized fraction are mildly inconsistent with the \textsc{late start/late end} model.  However, shifting the histories earlier by $\Delta z = 0.3$ removes these tensions.  }
    \label{fig:ion_history_lowz}
\end{figure}

Finally, we compare our reionization models to observational constraints on $x_{\rm HI}$ at $z \leq 6.5$ obtained using QSO spectra.  Figure~\ref{fig:ion_history_lowz} compares our models with constraints from Ly$\alpha$ forest dark pixels~\citep{McGreer2015,Jin2023}, dark gaps~\citep{Zhu2022}, QSO damping wings~\citep{Greig2024}, $P(< \tau_{\rm eff}^{50})$~\citep[][]{Choudhury2021,Gaikwad2023}, and Ly$\alpha$ forest damping wings~\citep{Zhu2024,Spina2024}.  The bold lines show the reionization histories in our two late-ending models, while the faded lines show these shifted to the right by $\Delta z = 0.3$, consistent with the discussion surrounding $P(< \tau_{\rm eff}^{50})$ in \S\ref{subsubsec:lya}.  We caution that most of these measurements are model-dependent to some degree, so showing them on the same plot may not constitute a fair comparison.  However, our goal here is to determine if measurements using different techniques and assumptions, {\it taken at face value}, display a clear preference for one of our scenarios.   

Most constraints are upper (lower) limits on the neutral (ionized) fraction.  Several of these are in mild tension with the \textsc{late start/late end model}, and most are consistent with the \textsc{early start/late end} case.  The $z = 5.5$ dark gap constraint from~\citealt{Zhu2022} and the recent QSO damping wing limits from~\citealt{Greig2024} disfavor the \textsc{late start/late end} model.  The recent {\it lower} limit on the neutral fraction from~\citealt{Zhu2024}, derived from Ly$\alpha$ forest damping wings at $z = 5.8$, disfavors an end to reionization early than this.  The same can be said of the~\citealt{Spina2024} forest damping wing measurement at $z = 5.6$, although their measurement actually prefers the \textsc{late start/late end} case.  An important caveat is that if reionization ends earlier by $\Delta z = 0.3$ (as hinted by $P(< \tau_{\rm eff}^{50})$ measurements), the tension with the \textsc{late start/late end} model disappears.  In fact, the neutral fraction in the shifted \textsc{early start/late end} model cannot be much lower without being in tension with the~\citealt{Zhu2024} damping wing limit.  We conclude that neutral fraction constraints at $z < 6.5$ mildly prefer the \textbf{\textsc{early start/late end}} model, but that relatively small, realistic shifts in the reionization history could change this conclusion.  

\subsection{Ly$\alpha$ Emitters at $z > 8$} \label{subsec:LAEs}

In this section, we study the Ly$\alpha$ transmission properties at $z \geq 8$ around massive halos that could host bright LAEs, such as GN-z11~\citep{Bunker2023}.  Our goal is to determine whether these observations prefer an early or late start to reionization.  \\

\subsubsection{Examples of IGM transmission at $z = 8$}
\label{subsec:Lyatrans}

\begin{figure}
    \centering
    \includegraphics[scale=0.25]{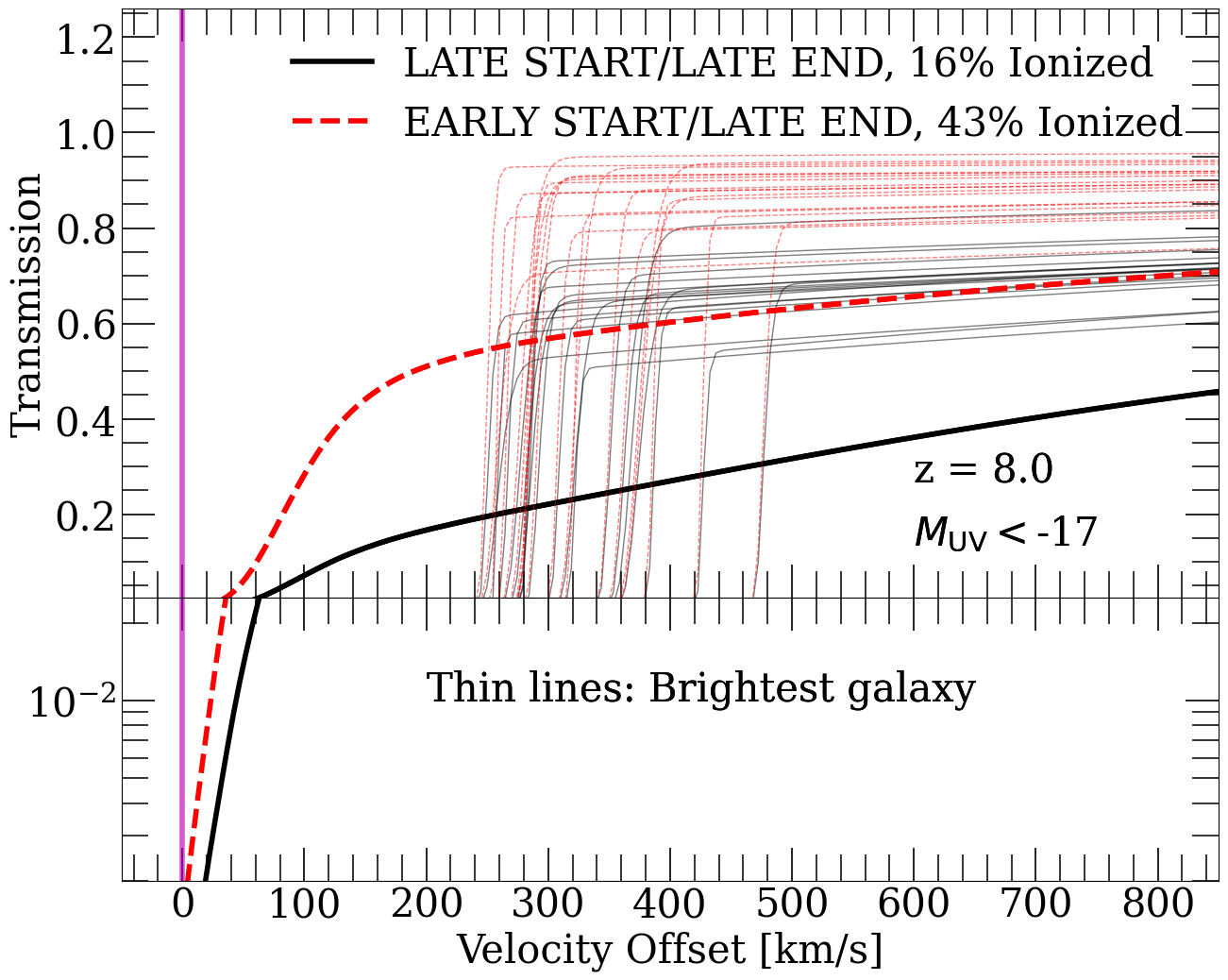}
    \caption{Mean IGM Ly$\alpha$ transmission vs. velocity offset around galaxies with $M_{\rm UV} < -17$ at $z = 8$ in our late-ending models.  The thick curves show the mean transmission profiles, which go to $0$ near systemic Ly$\alpha$ ($v_{\rm off} = 0$, magenta line), and rise at $v_{\rm off} > 0$, displaying a shape similar to the damping wing absorption profile.  The thin lines denote $20$ profiles for individual sightlines surrounding the brightest galaxy (most massive halo) in our box.  The \textsc{early start/late end} model has much more transmission on average than the \textsc{late start/late end} case, by a factor of $2$ or more at $v_{\rm off} < 500$ km/s.  This owes to its higher ionized fraction (see legend).  The sightlines surrounding the brightest galaxy display more transmission at $v_{\rm off} \gtrsim 200-400$ km/s than the average, and less at smaller $v_{\rm off}$ owing to large-scale inflows surrounding that object.  The higher transmission owes to the large, biased ionized bubble inhabited by the most massive halo in the box.  This illustrates that even in models with low average Ly$\alpha$ transmission, individual sightlines can still be transmissive, especially towards the brightest galaxies.  }
    \label{fig:mean_transmission}
\end{figure}

In Figure~\ref{fig:mean_transmission}, we illustrate how Ly$\alpha$ transmission surrounding bright galaxies differs in the \textsc{late start/late end} and \textsc{early start/late end} models at $z = 8$.  The solid curves show the IGM transmission ($T_{\rm IGM}$) vs. velocity offset ($v_{\rm off}$) on the red side of systemic averaged over halos with $M_{\rm UV} < -17$.  The vertical magenta line denotes systemic Ly$\alpha$, $v_{\rm off} = 0$.  $T_{\rm IGM}$ goes to $0$ at systemic, and at $v_{\rm off} > 0$ displays a shape similar to the characteristic damping-wing profile.  We see much higher Ly$\alpha$ transmission in the \textsc{early start/late end} case, owing to its much higher ionized fraction (indicated in the legend).  At $v_{\rm off} < 500$ km/s, $T_{\rm IGM}$ is a factor of $2$ or more above the \textsc{late start/late end} model.  The thin lines show individual transmission profiles for $20$ sightlines surrounding the brightest galaxy in the box.  These are much higher than the average at $v_{\rm off} \gtrsim 200-400$ km/s, and drop below the mean at smaller $v_{\rm off}$ due to fast in-flowing gas around this object (see discussion of inflows in \S\ref{subsec:lyatransmission}). The higher transmission at large $v_{\rm off}$ owes to this object occupying a larger ionized region than the average galaxy.

The higher $T_{\rm IGM}$ in the \textsc{early start/late end} model would seem to naturally explain the detection of bright galaxies hosting Ly$\alpha$ emission at $z \geq 8$.  However, it is important to note that even the \textsc{late start/late end} model displays some transmission, even with its $16\%$ ionized fraction, and this can be fairly high around the most biased objects (as the thin black lines show).  Indeed, even around this single halo there is significant sightline-to-sightline scatter.  This suggests that a statistical sample of observations is required to judge conclusively which model is preferred~\citep{Smith2022,Perez2023}.  It also suggests that some LAE detections at $z \geq 8$ could be explainable even if reionization starts relatively late.  It should be noted also that our finite box size ($200$ $h^{-1}$Mpc) may result in an under-estimate of average damping wing transmission around galaxies, as recently pointed out by~\cite{Keating2023}.

\subsubsection{Visibility of LAEs}
\label{subsubsec:LAE_visibility}

For an LAE with an intrinsic equivalent width EW$_{\rm int}$, and an average IGM transmission over the emitted line, $\langle T_{\rm IGM}\rangle_{\rm line}$, the observed equivalent width EW$_{\rm obs}$ is
\begin{equation}
    \label{eq:EWobs}
    {\rm EW}_{\rm obs} = \langle T_{\rm IGM}\rangle_{\rm line} {\rm EW}_{\rm int}
\end{equation}
An object is detectable if EW$_{\rm obs}$ is greater than some threshold ${\rm EW}_{\rm obs}^{\min}$.  This condition can be expressed as
\begin{equation}
    \label{eq:Tthresh}
    \frac{{\rm EW}_{\rm obs} }{{\rm EW}_{\rm int} }  = \langle T_{\rm IGM} \rangle_{\rm line} > \frac{{\rm EW}_{\rm obs}^{\min} }{{\rm EW}_{\rm int} } \equiv T_{\rm thresh}
\end{equation}
where we have defined $T_{\rm thresh}$ as the minimum IGM transmission that would make the LAE detectable.  To avoid assumptions about the intrinsic properties of the LAE population, we will parameterize our visibility calculations in terms of $T_{\rm thresh}$.  We will also adopt the simplification that $\langle T_{\rm IGM} \rangle_{\rm line}$ can be approximated by $T_{\rm IGM}$ at the $v_{\rm off}$ of the line's emission peak\footnote{This assumption is not true in general because $T_{\rm IGM}$ can vary significantly over the width of the emission line.  }.  This allows us to parameterize LAE visibility in the $(T_{\rm thresh}, v_{\rm off})$ parameter space.  In this section, we calculate LAE visibility statistics at $z = 8$, $9$, $10$, and $11$.  

\begin{figure*}
    \centering
    \includegraphics[scale=0.24]{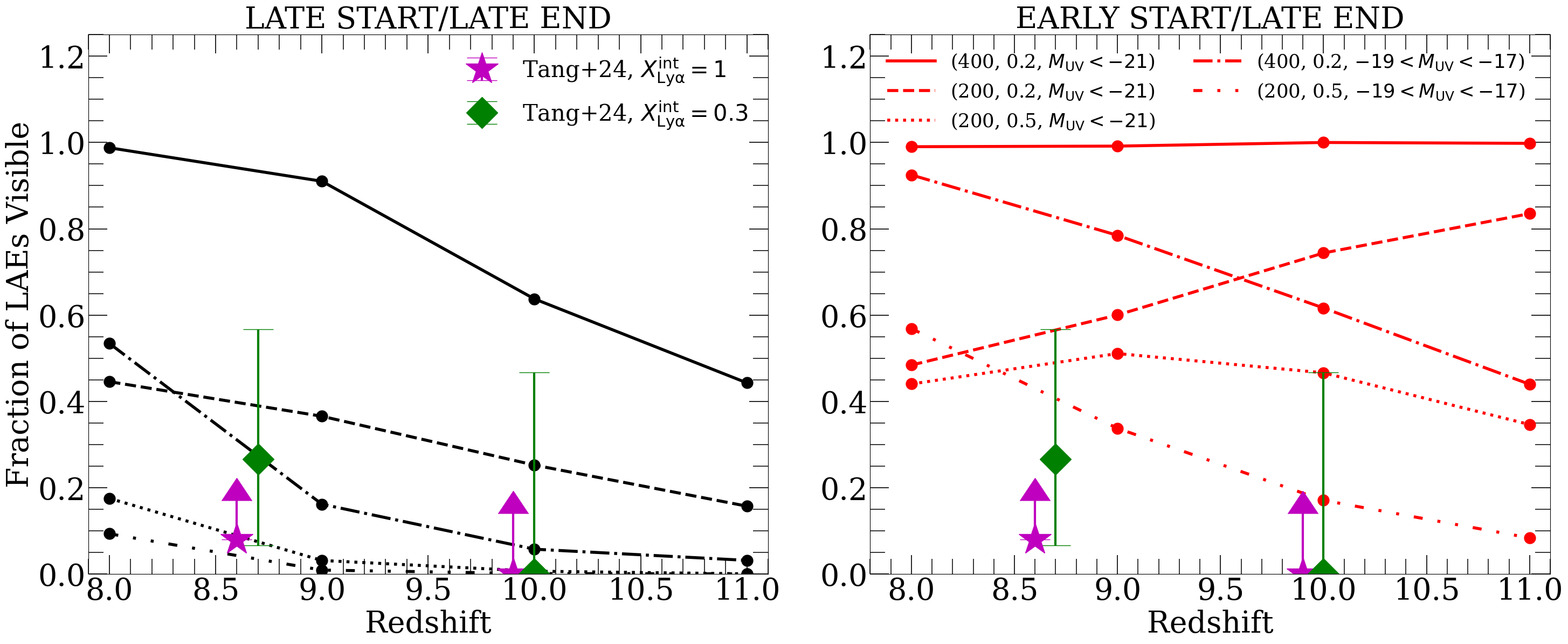}
    \caption{Evolution of LAE visibility with redshift for several combinations of $v_{\rm off}$ (in km/s), $T_{\rm thresh}$, and $M_{\rm UV}$ range - denoted as $(v_{\rm off}, T_{\rm thresh}, M_{\rm UV}\text{ range})$ in the legend in the right panel.  The left and right panels show the \textsc{late start/late end} and \textsc{early start/late end} models, respectively.  The solid, dashed, and dotted  curves show visibilities for bright galaxies with $M_{\rm UV} < -21$ - these three curves have $(v_{\rm off}, T_{\rm thresh}) = (400\text{ km/s}, 0.2)$, $(200\text{ km/s}, 0.2)$, and $(200 \text{ km/s}, 0.5)$, respectively.  The dot-dashed and double-dot dashed curves show faint galaxies ($-19 < M_{\rm UV} < -17$), with $(v_{\rm off}, T_{\rm thresh}) = (400\text{ km/s}, 0.2)$ and $(200\text{ km/s}, 0.5)$, respectively.  Most choices of these parameters yield $\approx 50-100\%$ visibility in the \textsc{early start/late end} model, the only exception being faint LAEs with low $v_{\rm off}$ and high $T_{\rm thresh}$ (double-dot dashed).  By contrast, the \textsc{late start/late end model} displays a wide range of expected LAE visibility statistics for different types of LAEs.   In the left panel, we show recent measurements of the LAE fraction from~\citealt{Tang2024b}.  The magenta points show the fraction of galaxies observed to have Ly$\alpha$ EW $> 25\text{\AA}$.  Interpreting this as a {\it fraction of LAEs} that are visible means assuming $X_{\rm Ly\alpha}^{\rm int} = 1$, so these points are shown as lower limits.  The green points assume $X_{\rm Ly\alpha}^{\rm int} = 0.3$, as measured at $z = 5$ by~\citealt{Tang2024a}.  The low measured visibility fractions are consistent with range of visibility statistics found in the \textsc{late start/late end} case.  However, in the \textsc{early start/late end} model, only the most ``restrictive'' parameter combination that we show - $T_{\rm thresh} = 0.5$, $v_{\rm off} = 200$ km/s, and $-17 < M_{\rm UV} < -19$ displays similar statistics.  }
    \label{fig:transmission_evolution}
\end{figure*}

Recently,~\citealt{Asthana2024} studied the distribution of ionized bubble sizes in reionization models similar to ours.  Generally, it is expected that galaxies must inhabit an ionized bubble of radius $\gtrsim 1$ pMpc to guarantee a high level of Ly$\alpha$ transmission on the red side of systemic~\citep{Weinberger2018,Mason2020}.  They found that a model similar to our \textsc{early start/late end} scenario is required to produce a significant number of such ionized bubbles $8 \leq z \leq 10$.  In Figure~\ref{fig:transmission_evolution}, we perform a similar analysis using our visibility calculations.  We show the fraction of LAEs with $T_{\rm IGM} > T_{\rm thresh}$ vs. redshift for several choices of $T_{\rm thresh}$, $v_{\rm off}$, and UV magnitude range (faint vs. bright galaxies).  The caption gives these parameter combinations for each curve as a brightness ($M_{\rm UV}$) and a combination of $T_{\rm thresh}$ and $v_{\rm off}$.  Visibility fractions increase with decreasing $T_{\rm thresh}$ and increasing $v_{\rm off}$.  The latter is true because $T_{\rm IGM}$ increases with $v_{\rm off}$ as the damping wing opacity decreases (Figure~\ref{fig:mean_transmission}).  Visibility is also higher for brighter galaxies, which inhabit the largest ionized bubbles.  The left and right panels show results for the \textsc{late start/late end} and \textsc{early start/late end} models, respectively.  

The solid curves show visibility for bright ($M_{\rm UV} < -21$) LAEs with large velocity offsets ($400$ km/s) and low visibility thresholds ($T_{\rm thresh} = 0.2$).  Such objects are visible nearly $100\%$ of the time in the \textsc{early start/late end} case, and $40\%$ of the time in the \textsc{late start/late end} model even at $z = 11$.  Reducing $v_{\rm off}$ to $200$ km/s (dashed curves) deceases visibility, especially in the \textsc{late start/late end} case, but even then $20-40\%$ of LAEs are visible at $8 < z < 11$.  In the \textsc{early start/late end} case, the visibility fraction counter-intuitively {\it increases} with redshift.  This is because the evolution in visibility is not being driven by the neutral fraction, but by inflows surrounding massive halos.  At higher redshifts, brighter objects are found in less massive halos, which are surrounded by smaller inflows.  This leads to increased transmission at $v_{\rm off} = 200$ km/s~\citep{Park2021}.  

The dotted curves show that increasing $T_{\rm thresh}$ from $0.2$ to $0.5$ (for $M_{\rm UV} < -21$ and $v_{\rm off} = 200$ km/s) has a substantial effect on visibility.  In the \textsc{late start/late end} model, $< 20\%$ of LAEs are visible at $z = 8$, and this drops to near-$0$ at $z \geq 9$.  However, in the \textsc{early start/late end} case, $40-50\%$ of such objects are visible across this redshift range.  The dot dashed curve considers faint ($-19 < M_{\rm UV} < -17$) galaxies with high $v_{\rm off} = 400$ km/s and low $T_{\rm thresh} = 0.2$.  These objects are visible $50-90\%$ of the time in the \textsc{early start/late end} model, but $< 20\%$ of the time in the \textsc{late start/late end} case.  Finally, the double-dot dashed curves also show faint galaxies, but with $v_{\rm off} = 200$ km/s and $T_{\rm thresh} = 0.5$, a parameter combination that minimizes LAE visibility.  In the \textsc{late start/late end} case, fewer than $10\%$ of such objects are visible at any redshift, while in the \textsc{early start/late end} model, $17\%$ ($8\%$) of such objects are visible at $z = 10$ ($11$), and over half are visible at $z = 8$.  

In the left panel, we show recent measurements of the fraction of galaxies hosting Ly$\alpha$ emitters, (the Ly$\alpha$ fraction, $X_{\rm Ly\alpha}^{\rm obs}$), at $z = 8.7$ and $z = 10$ from~\citealt{Tang2024b}.  The purple points show $X_{\rm Ly\alpha}^{\rm obs}$ as measured from that work.  Comparing these points directly with the {\it fraction of LAEs} that are visible assumes that the intrinsic fraction of galaxies hosting LAEs is unity - that is, $X_{\rm Ly\alpha}^{\rm int} = 1$.  For this reason, we display these points as lower limits on $X_{\rm Ly\alpha}^{\rm obs}/X_{\rm Ly\alpha}^{\rm int}$.  The green points show $X_{\rm Ly\alpha}^{\rm obs}/X_{\rm Ly\alpha}^{\rm int}$ assuming $X_{\rm Ly\alpha}^{\rm int} = 0.3$, the Ly$\alpha$ fraction measured at $z = 5$ by~\citealt{Tang2024a}, which we do not show as limits.  The curves in the left panel show wide spread that is broadly consistent with the measured visibilities.  However, in the right panel, all the curves are on the high end of the measurements (except for the double-dot dashed curve).  At face value, these findings indicate that the observed visibility of LAEs is too low for the \textsc{early start/late end} scenario, instead preferring the \textbf{\textsc{late start/late end}} model.  

For the global LAE visibility fraction to evolve consistently with measurements from~\citealt{Tang2024b} in the \textsc{early start/late end} model, the double-dot dashed curve in Figure~\ref{fig:transmission_evolution} would have to characterize the bulk of the population.  These are faint LAEs with emission at low $v_{\rm off}$ that require high IGM transmission to observe.  While it is true that faint LAEs tend to have low $v_{\rm off}$~\citep{Mason2018a}, they also tend to have fairly high EW$_{\rm intr}$~\citep{Dijkstra2012,Tang2024a}.  A majority of the LAEs observed at $z = 5$ by~\citealt[][]{Tang2024a} in that $M_{\rm UV}$ range have EW$_{\rm intr} > 50\text{\AA}$, and about half have EW$_{\rm intr} > 100\text{\AA}$.  With the visibility threshold of EW$_{\rm obs}^{\min} = 25\text{\AA}$ used in~\citealt{Tang2024b}, this would imply $T_{\rm thresh} < 0.5$ for the majority of faint objects, and $T_{\rm thresh} < 0.25$ for half of them.  Moreover, a significant fraction of the faint objects observed at $z = 5-6$ in~\citealt[][]{Tang2024a} have $v_{\rm off} > 200$ km/s.   Brighter galaxies in their sample generally have smaller EW$_{\rm intr}$, but these also tend to have higher $v_{\rm off}$~\citep[see also][]{Mason2018a} and inhabit larger bubbles, such that they would remain visible in the \textsc{early start/late end} model even if they required higher $T_{\rm thresh}$ to detect.  By contrast, in the \textsc{late start/late end} model, LAEs with a wide range of properties have visibilities consistent with the~\citealt[][]{Tang2024b} measurements.   

\subsubsection{Does GN-z11 require an early start?}
\label{subsec:gnz11}

GN-z11 is the highest-redshift LAE detected to date, at $z = 10.6$.  It has a broad Ly$\alpha$ emission feature centered at $v_{\rm off} \approx 550$ km/s, a full-width at half maximum (FWHM) of $\Delta v \approx 400$ km/s, and an observed EW of $18\text{\AA}$.  Using a Bayesian analysis based on reionization simulations and an empirically derived model for the intrinsic EW distribution of LAEs from~\citealt{Mason2018a},~\citealt{Bruton2023} inferred that the IGM must be at least $12\%$ ionized at $2 \sigma$ confidence (yellow point in Figure~\ref{fig:ion_history_LAE}).  This constraint, at face value, clearly favors the \textsc{early start/late end} model (see Figure~\ref{fig:ion_history_LAE} in the next section).  Here, we consider whether the observed properties of GN-z11 require reionization to start early. 

\begin{figure}
    \centering
    \includegraphics[scale=0.25]{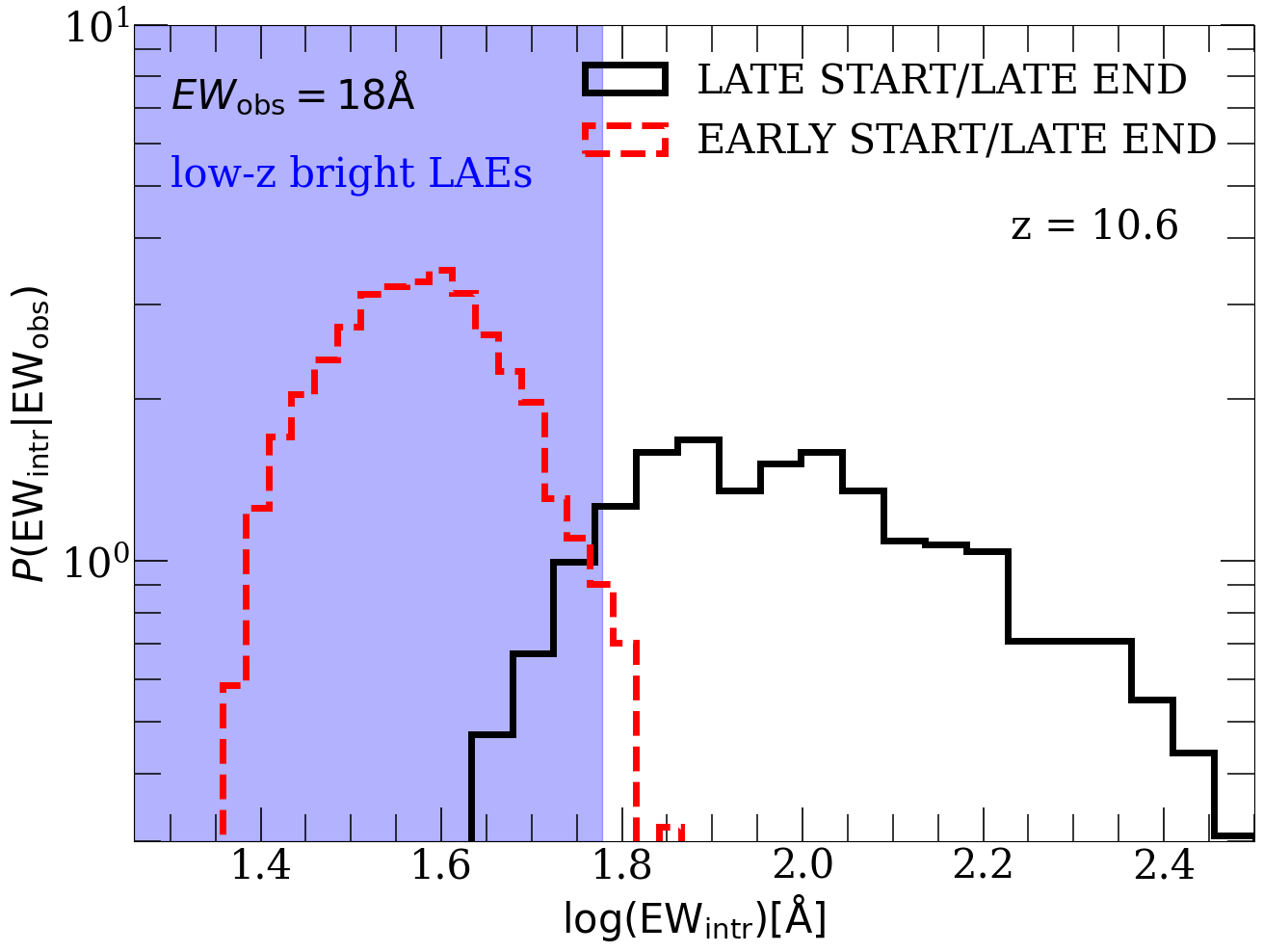}
    \caption{Distribution of EW$_{\rm intr}$ required to produce observed an LAE with the properties of GN-z11 at $z = 10.6$ (see text for details).  The shaded region denotes EW$_{\rm intr} < 60\text{\AA}$, the range observed for similarly bright galaxies at low redshift (see text for references).  We see that nearly all sightlines to bright galaxies in the \textsc{early start/late end} model have EW$_{\rm intr}$ within this shaded region, and with $20\%$ ($95\%)$ of sightlines requiring EW$_{\rm intr} < 30 \text{\AA}$ ($60\text{\AA}$).  By contrast, the \textsc{late start/late end} model has a wider distribution centered at higher EW$_{\rm intr}$.  About $12\%$ of sightlines in this model require EW$_{\rm intr} < 60\text{\AA}$ though, suggesting that a particularly bright LAE with relatively high EW viewed along a sightline with properties that occur $\approx 10\%$ of the time could produce an observation like GN-z11 in the \textsc{late start/late end} scenario.  }
    \label{fig:GNz11}
\end{figure}

We can estimate the EW$_{\rm intr}$ required to produce the observed GN-z11 emission line as follows.  First, we model the intrinsic line as a Gaussian with some central velocity $v_{\rm off}^{\rm intr}$, FWHM $\Delta v^{\rm intr}$, and amplitude $A$.  Then, the observed emission profile for a given sightline is given by the intrinsic profile multiplied by $T_{\rm IGM}(v_{\rm off})$.  We also model the observed line as a Gaussian, with parameters given in the previous paragraph, and the continuum and normalization chosen to give the observed EW.  For a sample of $\sim 2000$ sightlines surrounding $-22 < M_{\rm UV} < -21$ galaxies, we fit for the parameters of the intrinsic line that, after attenuation by the IGM, gives a best fit to the observed line.  The distribution of EW$_{\rm intr}$ recovered with this procedure, $P({\rm EW}_{\rm int} | {\rm EW}_{\rm obs})$, at $z = 10.6$ is shown in Figure~\ref{fig:GNz11} for our late-ending models.  The shaded region denotes the range of EW observed in similarly bright galaxies at lower redshifts~\citealt{Endsley2022,Tang2023,Saxena2024,Tang2024a}, the highest of which is $\approx 60\text{\text{\AA}}$ (Fig. 1 of~\citealt{Tang2024b}). 

We see a stark contrast between $P({\rm EW}_{\rm int} | {\rm EW}_{\rm obs})$ in our models.  Nearly all the sightlines in the \textsc{early start/late end} model require EW$_{\rm intr} < 60\text{\AA}$, and about $20\%$ require EW$_{\rm intr} < 30\text{\AA}$.   This suggests that bright LAEs with EWs on the high end of the observed distribution will produce a GN-z11-like observation most of the time in this scenario.  In the \textsc{late start/late end} model, the distribution is much wider and shifted to much higher EW$_{\rm intr}$.  Only $12\%$ of sightlines allow for EW$_{\rm intr} < 60\text{\AA}$, and none of them allow $< 30\text{\AA}$.  So, although objects such as GN-z11 are expected to be fairly rare in the \textsc{late end/late start} scenario, they would not be impossible to find.  Note that the two $M_{\rm UV} < -21$ LAEs observed in~\citealt{Tang2024b} with the highest EWs ($\approx 30$ and $60\text{\AA}$) were both detected in H$\alpha$.  Based off a clear detection of H$\gamma$ emission in GN-z11,~\citealt{Bunker2023} estimated that it should have strong H$\alpha$ emission.  As such, we can conclude that the detection of GN-z11 does not rule out the \textsc{late start/late end} scenario.  However, if forthcoming observations reveal similar objects to be ubiquitous at $z > 10$, it would be strong evidence in favor of something similar to the \textsc{early start/late end} model.    

\subsubsection{Constraints on $x_{\rm HI}$ with galaxies at $z \geq 6.5$}
\label{subsec:neutral_frac_z6.5}

\begin{figure}
    \centering
    \includegraphics[scale=0.19]{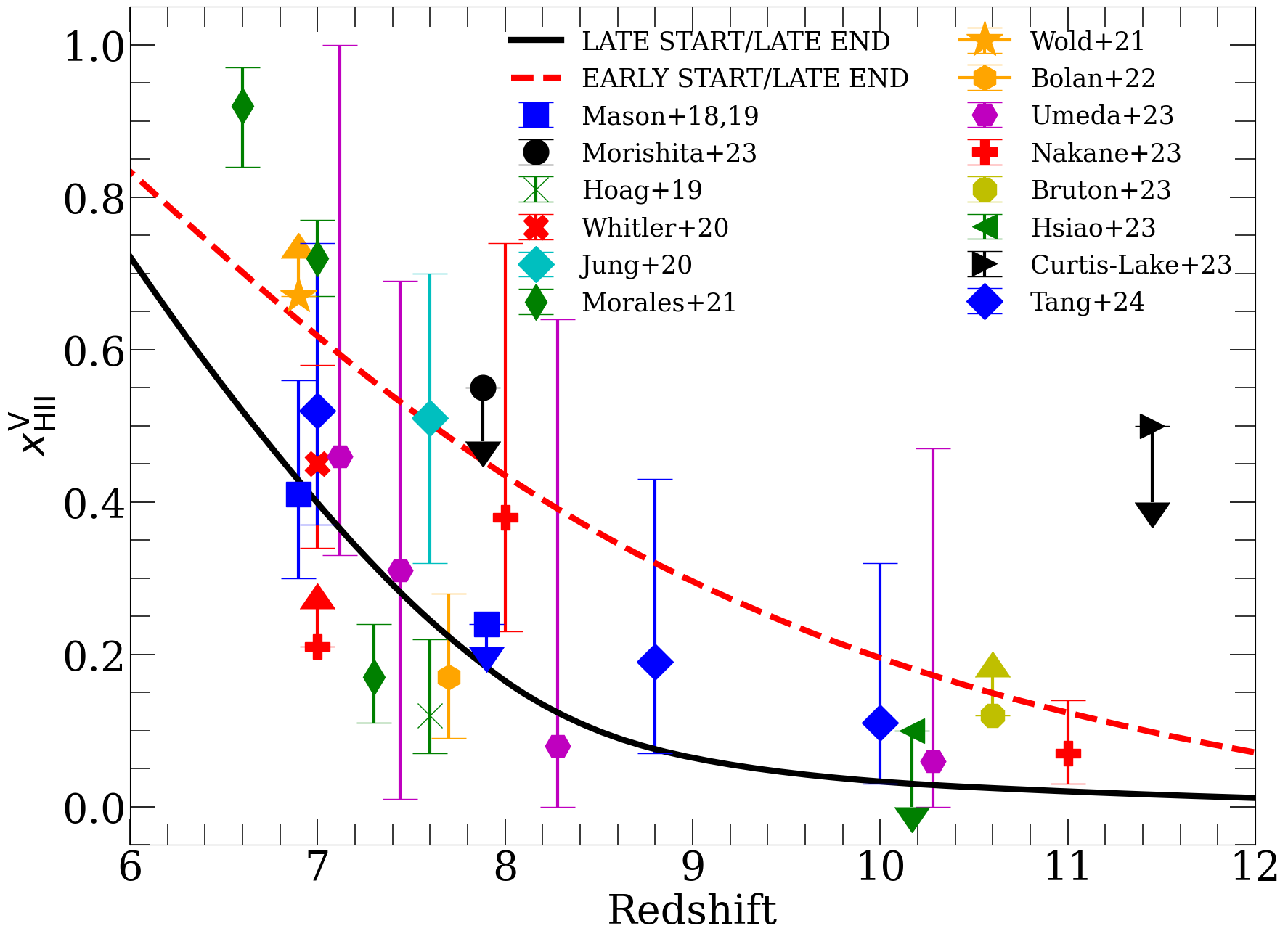}
    \caption{Summary of neutral fraction measurements in the literature based on either Ly$\alpha$ emission or Ly$\alpha$ damping wings in galaxy spectra.  Most of the measurements are at $7 < z < 8$, and taken together they do not strongly prefer either late-ending model over the other.  There is a dearth of constraints at $8.5 < z < 10$ (the exception being the~\citealt{Tang2024b} point at $z \sim 8.7$), and those at $z > 10$ do not show a clear consensus either.  The only measurement that shows a strong preference for the \textsc{early start/late end} model is the constraint from~\citealt{Bruton2023}, based on GN-z11 (see \S\ref{subsec:gnz11}).  Measurements are from~\citealt{Mason2018a,Mason2019,Hoag2019,Whitler2020,Jung2020,Morales2021,Bolan2022,Wold2022,Umeda2023,Morishita2023,Nakane2023,Bruton2023,Hsiao2023,Curtis-Lake2023,Tang2024b}.  }
    \label{fig:ion_history_LAE}
\end{figure}

To conclude our discussion of LAEs, we look at measurements of $x_{\rm HI}$ from observations of galaxies at $z \geq 6.5$.  These include constraints from the statistics of LAE detections, and those based on Ly$\alpha$ damping wing absorption in galaxy spectra.  Figure~\ref{fig:ion_history_LAE} shows a collection of these measurements and limits compared to our late-ending reionization models in the same format as Figure~\ref{fig:ion_history_lowz} (with references in the caption).  These constraints are all model-dependent to some degree, so showing them on the same plot may not constitute a fair comparison.  Our goal here is to illustrate the diversity of constraints obtained across multiple observations and inference techniques.  

Unlike in Figure~\ref{fig:ion_history_lowz}, we see no clear preference for either scenario.  Indeed, at $z < 8$, several constraints prefer each of the models, while some have error bars too large to distinguish them.  The only consensus that these constraints give collectively is that reionization is in progress at $7 < z < 8$.  At $z > 10$, all the constraints are based on damping wings except that of~\citealt{Bruton2023}, which is based on the detection of GN-z11.  There is no clear consensus between these constraints either.  There is a notable dearth of constraints at $8.5 < z < 10$, the redshift range where the two models differ the most.  The only exception is the~\citealt{Tang2024b} point at $z \sim 8.7$, which falls exactly between our models but has large error bars.  It seems clear from this comparison that constraints on $x_{\rm HI}$ from high-redshift galaxies do not, at present, display a clear preference for either a late or early start to reionization.  Note that we have included here only constraints on $x_{\rm HI}$ derived from galaxies, since that is our main focus on this section. However, a number of additional measurements have been made using QSO damping wing absorption, especially at $6.5 < z < 8$~\citep[e.g.][]{Davies2018,Yang2020a,Durovcikova2024}.  These measurements do not display a clear collective preference for a late or early start either.  

\subsection{Patchy kSZ from reionization} \label{subsec:cmb}

\begin{figure*}
    \centering
    \includegraphics[scale=0.35]{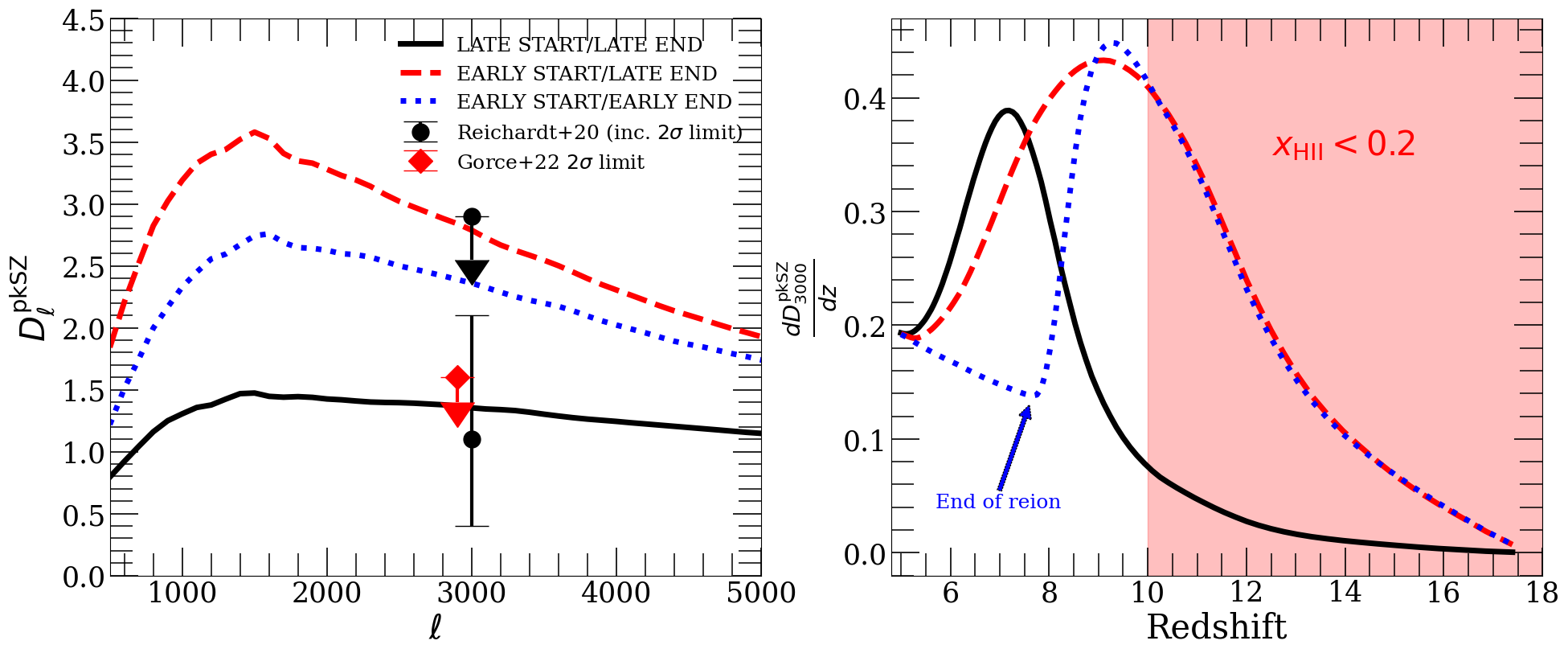}
    \caption{Patchy kSZ from reionization for all three of our model.  {\bf Left:} patchy kSZ power vs. $\ell$, compared to the recent measurement and $2\sigma$ upper limit at $\ell = 3000$ from~\citealt{Reichardt2020}.  We also include the $2\sigma$ upper limit from the re-analysis of the SPT data by~\citealt{Gorce2022} (offset for readability), which disfavors the \textsc{early start/late end} model.  Only the \textsc{late start/late end} model falls within the $1\sigma$ errors, and the \textsc{early start/late end} case has the most power and lies close to the $2 \sigma$ upper limit from~\citealt{Reichardt2020}.  {\bf Right:} differential contribution to the total power at $\ell = 3000$ as a function of redshift.  The shaded red region shows the range where the universe is less than $20\%$ ionized in the \textsc{early start/late end} model, from which nearly half the $\ell = 3000$ power originates.  The \textsc{early start/early end} model has slightly less power than the \textsc{early start/late end} case because it ends reionization earlier and thus has a shorter duration (see annotation).  }
    \label{fig:pksz}
\end{figure*}

In this section, we will turn again to the CMB to help distinguish our reionization models.  We display the patchy kSZ power spectra for all three models
(see \S\ref{subsec:model_pksz}) in the left panel of Fig. \ref{fig:pksz}, along with the $1\sigma$ error bar and 95$\%$ confidence upper limits from \citealt{Reichardt2020}.  We see that the \textsc{late start/late end} model alone lies within the $1\sigma$ of the SPT measurement.  The other two both fall outside this range but still within the 95$\%$ confidence upper limit, with the \textsc{early start/late end} case coming closest to the upper limit\footnote{We note that any measurement of the pkSZ involves assumptions about the late time, homogeneous kSZ.  The methods in \citealt{Reichardt2020} assume an end to reionization at $z_{\rm end}$ = 5.5, whereas our simulations (and our pkSZ contributions) continue until $z_{\rm end} \sim$ 5.  Correcting for this would bring the measurement up slightly ($\sim$ 0.1 $\mu$K$^2$), but not enough to qualitatively affect our conclusions.  }.  We also include the revised $2 \sigma$ upper limit from~\citealt{Gorce2022}, which is somewhat lower than the SPT result and favors the \textsc{late start/late end} model even more.  

To gain intuition for the origin of the differences in pkSZ power, we plot the differential contribution to $D_{\ell=3000}$ per $z$ in the right panel.  Both early-starting models begin contributing power as soon as reionization starts at $z = 18$, as ionized bubbles form and grow to sufficient scales.  The two begin diverging at $z \sim 10$, as the \textsc{early start/early end} case finishes reionization, causing the pkSZ power at $\ell=3000$ to drop abruptly at $z = 8$ (see annotation).  This fall-off in power corresponds to the disappearance of large-scale ionization fluctuations, at which point the features in the kSZ signal on these scales are set by fluctuations in density and velocity only.  In contrast, ionization fluctuations persist longer in the \textsc{early start/late end} model, and so continue to contribute power to the pkSZ signal at $\ell = 3000$ at $z < 8$.  

In the \textsc{late start/late end} case, reionization begins much later but still ends at $z = 5$, which makes the peak in $dD_{3000}^{\rm pkSZ}/dz$ narrower.  The shaded red region shows that nearly half the power in the \textsc{early start/late end} case arises at $z > 10$ when the ionized fraction is $< 20\%$ in that model.  In the \textsc{late start/late end} case, reionization is just starting around $z = 10$.  Thus, we see that the pkSZ is highly sensitive to reionization's duration, and particularly its early stages~\citep{Battaglia2012,Chen2022}.  We see also that although the~\citealt{Reichardt2020} measurement does not rule out the \textsc{early start/late end} model at $2\sigma$, it clearly prefers the \textbf{\textsc{late start/late end}} case.  This finding is consistent with the recent limits on the duration of reionization from~\citealt{Raghunathan2024} using data from SPT and the {\it Herschel}-SPIRE experiment.  They found that $\Delta z_{50}$, the difference between redshifts at $25\%$ and $75\%$ ionized fractions, is $< 4.5$ at $95\%$ confidence\footnote{One caveat is that their constraint assumes that reionization ends by $z = 6$, whereas in our models it completes at $z \approx 5$.  Their constraint would loosen by $dz \sim 1$ if $z = 5$ were assumed.  } - our \textsc{early start/late end} model has $\Delta z_{50} = 3.1$.  

\section{Discussion} \label{sec:disc}

\subsection{``Face-value'' interpretations of the data}

 \begin{table*}
        \Large
        \centering
        \hspace{-2.8cm}\begin{tabular}{|c|c|c|c|}
                \hline \hline
            \bf{Category} & {\bf Observable} & {\bf Late Start} & {\bf Early Start}\\
             \hline \hline
             \textcolor{blue}{CMB} & \textcolor{blue}{$\tau_{\rm es}$}  & No Preference & No Preference\\
             & \textcolor{blue}{Patchy kSZ} & \textcolor{olive}{Preferred} & \textcolor{orange}{Not preferred} \\
             \hline
              \textcolor{red}{High-$z$ Galaxies} & \textcolor{red}{UVLF/$\xi_{\rm ion}$}  &  \textcolor{olive}{Preferred} & \textcolor{orange}{Not preferred}\\
                & \textcolor{red}{LAEs at $z > 8$}  & \textcolor{olive}{Preferred} & \textcolor{orange}{Not preferred}\\
                & \textcolor{red}{$x_{\rm HI}$($z > 6.5$)} & No Preference & No Preference\\
               \hline
              \textcolor{magenta}{$z < 6.5$ QSOs} & \textcolor{magenta}{$\langle F_{\rm Ly\alpha} \rangle$}  & No Preference & No Preference\\
             & \textcolor{magenta}{$P(< \tau_{\rm eff}^{50})$} & \textcolor{orange}{Not preferred} & \textcolor{olive}{Preferred}\\
             & \textcolor{magenta}{Mean Free Path} & \textcolor{olive}{Preferred} & \textcolor{orange}{Not preferred}\\
             & \textcolor{magenta}{Thermal History}  & \textcolor{orange}{Not preferred} & \textcolor{olive}{Preferred}\\
             & \textcolor{magenta}{$x_{\rm HI}$($z < 6.5$)} & \textcolor{orange}{Not preferred} & \textcolor{olive}{Preferred}\\
            \hline \hline
                     Final Score & All Data & \textcolor{olive}{Preferred} & \textcolor{orange}{Not preferred}\\% & \textcolor{red}{Disfavored}\\
                     \hline \hline
        \end{tabular}
        \caption{Summary of the ``face value'' preferences of each observable we studied for a late or early start to reionization.  The left-most column gives the category for each observable - whether it derives from the CMB (blue), high-redshift galaxies (red), or $z < 6.5$ QSOs (magenta).  The second column lists the observables, grouped by category.  The right two columns give the preference of each probe for a late or early start to reionization.  Both columns list ``no preference'' if an observable does not clearly lean towards either scenario.  Importantly, neither scenario is ``confirmed'' or ``ruled out'' by any observables.  We see that not all the probes prefer the same scenario.  Three of the observables derived from QSO spectra ($P(<\tau_{\rm eff}^{50})$, $T_0$, and $x_{\rm HI}(z < 6.5)$, see \S\ref{subsec:QSO}) prefer an early start, while the remaining probes all either prefer a late start or cannot distinguish the two scenarios.  All three categories, which use largely independent data sets with vastly different analysis techniques, have at least one probe that prefers a late start.  However, all the probes that prefer an early start derive from the same type of data (QSO spectra).  These in particular are challenging to interpret due to uncertainties in modeling the $z < 6.5$ reionization history and connecting this to what happens at higher redshifts.  As such, we conclude that observations display a mild preference for the \textbf{\textsc{late start/late end}} scenario.  }
        \label{tab:summary}
    \end{table*}

In the previous sections, we studied how the properties of our three models compare to a broad range of observables.  These include measurements of the UVLF and $\xi_{\rm ion}$ from JWST (\S\ref{sec:JWST}), inferences from the spectra of high-redshift QSOs (\S\ref{subsec:QSO}), Ly$\alpha$ transmission from $z > 8$ galaxies (\S\ref{subsec:LAEs}), and constraints from the CMB (\S\ref{subsec:cmb}).  We concluded in \S\ref{subsec:QSO} that measurements of the Ly$\alpha$ forest at $z \leq 6$ strongly disfavor the \textsc{early start/early end} scenario.  However, the evolution of the $\langle F_{\rm Ly\alpha} \rangle$ alone could not distinguish between an early and late start to reionization.  Most of the other observables we studied individually displayed a preference for one or the other, but none could conclusively rule out either\footnote{We note that these preferences were determined primarily qualitative - based on ``chi-by-eye'' comparisons of models and observations, and as such they represent somewhat qualitative results.  Still, they are useful for gauging the direction that each data set will likely to push constraints on reionization if included in a quantitative analysis~\citep[e.g.][]{Qin2021,Nikolic2023}.  }.  This motivates the key question in this work: when taken together, what story do these observables tell about reionization's early stages?  Do they seem to push us in the direction of a late or an early start to reionization, and do they all agree?  Indeed, a major goal of the field is to synergize the constraining power of many observations to constrain reionization, and qualitative analyses like the one presented here can help chart the path for more detailed studies.  

We summarize our findings in Table~\ref{tab:summary}.  The left-most column lists the ``categories'' of observables that we studied - the CMB (blue), high-redshift galaxies (red), and $z < 6.5$ QSOs (magenta).  The second column from the left lists each of the observables, and the remaining two columns denote whether each observable prefers a late or early start to reionization.  We find that $\tau_{\rm es}$, $\langle F_{\rm Ly\alpha}\rangle$, and measurements of $x_{\rm HI}$ at $z > 6.5$ display no preference for either case.  JWST observations of the UVLF/$\xi_{\rm ion}$, the MFP, LAE visibility at $z > 8$, and the SPT pkSZ measurement prefer the \textsc{late start/late end} case.  By contrast, the Ly$\alpha$ forest $P(<\tau_{\rm eff}^{50})$, the IGM thermal history, and $x_{\rm HI}$ measurements at $z < 6.5$ prefer the \textsc{early start/late end} model.  

We see that not all these observables prefer the same scenario.  This suggests a possible lack of concordance between different data sets with respect to reionization's early stages and/or systematic deficiencies arising from inadequacies in theoretical modeling.  We note, however, that all three observables that favor the \textsc{early start/late end} model are based QSO spectra at $z \leq 6.5$.  Indeed, nearly all the data associated with these three observables (with the exception of the QSO damping wings) arises, directly or indirectly, from the Ly$\alpha$ forest, and thus cannot be treated as fully independent\footnote{Indeed, much of the data comes from the same QSO survey, XQR-30~\citep{DOdorico2023}.  }.  As we explained in \S\ref{subsec:QSO}, the conclusions we drew from these probes are sensitive to our modeling choices, which are necessary to link the late stages of reionization (which the forest probes directly) to its early stages.  By contrast, the observables that support a late start are derived from different data sets using vastly different techniques, and at least one probe in every category prefers a late start.  As such, we judge that that the \textbf{\textsc{late start/late end}} model is mildly preferred (overall) by observations.  

This seeming lack of consensus between different observables has several possible resolutions.  Perhaps the most straightforward is that existing observations lack the accuracy and/or precision to achieve a unanimous consensus about reionization's early stages.  Indeed, nearly all the observables considered here still have large uncertainties.  Theoretical modeling uncertainties, required to interpret the data, can similarly affect these conclusions.  As mentioned earlier, the QSO-based observables that seem to support an early start probe only reionization's end stages, requiring a model to infer the early history.  There are also several key modeling uncertainties and challenges associated with reionization, some of which are discussed above. These include the clustering and ionizing properties of the sources, absorption by the ionized IGM, and the computational challenge of surveying reionization's large parameter space with accurate models.  As a result, we do not interpret the seeming lack of agreement between observables seen here as a statistically significant tension.  These issues will continue to improve with time as more (and better) observational data is acquired and reionization models become faster and more accurate.  

However, a more concerning possibility is that forthcoming observations and rigorous theoretical analysis will reveal a clear tension between observables with respect to reionization's early stages.  In this scenario, there is danger of reaching a pre-mature conclusion that reionization has been solved with high confidence. This is because tensions between different data sets can lead to the spurious conclusion that the few models that are statistically compatible with all observations must be the correct ones.  Indeed, this is the reason we did not attempt to identify such a model in this work, and instead focused on the qualitative preferences displayed by each observable for different representative scenarios.  Forthcoming efforts should be aware of this potential pitfall, and take care to understand the effects of individual data sets on joint constraints.  

\subsection{Forthcoming observational prospects}
\label{subsec:obs_prospects}

In this section, we will briefly discuss prospects for future observations that would help strengthen the constraining power of some of the probes discussed here.  The first is to continue improving constraints on the UVLF and $\xi_{\rm ion}$, particularly for faint galaxies.  \citealt{Atek2023} demonstrated that $\xi_{\rm ion}$ could be measured reliably for faint ($-17 < M_{\rm UV} < -15$), lensed galaxies during reionization.  Such studies, together with efforts to directly constrain the faint end of the UVLF, will be crucial for determining the redshift evolution of these quantities and whether there is a fall-off in ionizing output for the faintest galaxies (see bottom panel of Figure~\ref{fig:alternatives}).  Continued efforts to understand how $f_{\rm esc}$ correlates with galaxy properties at low redshift, such as the Low-redshift Lyman Continuum Survey (LzLCS,~\citealt{Chisholm2022,Flury2022,Jaskot2024a,Jaskot2024b}) will also be crucial for placing reasonable limits on the evolution of $f_{\rm esc}$ (see also e.g.~\citealt{Smith2020,Pahl2021,Wang2023}).  

There is also further progress to be made with QSO-based observations at $z \sim 6$.  Improved constraints on the mean free path and IGM thermal history may help distinguish an early vs. late start, as Figure~\ref{fig:qso_summary_2} shows.  Forthcoming observations with Euclid~\citep{Atek2024} will dramatically increase the number of known quasars at these redshifts, allowing for spectroscopic follow-up that will improve statistical uncertainties on both sets of measurements.  Efforts to measure the relationship between Ly$\alpha$ forest opacity and galaxy density~\citep{Christenson2021,Ishimoto2022,Christenson2023,Kashino2023} may also help tighten constraints on the reionization history at $z < 6.5$ \citep[][]{Garaldi2022,Gangolli2024}.  

Further observations with JWST will improve the statistics of $z > 8$ galaxies displaying significant Ly$\alpha$ emission.  They will also yield constraints on $x_{\rm HI}$ from Ly$\alpha$ damping wing absorption at redshifts where very few bright quasars are available.  Forthcoming observations with the Nancy Grace Roman telescope~\citep{Wold2024} will also reveal bright LAEs over a much wider area than JWST, enabling improved constraints on the early reionization history~\citep{Perez2023}. 

Forthcoming improvements on CMB constraints from multiple experiments, including the Atacama Cosmology Telescope~\citep[ACT,][]{Hlozek2012}, SPT~\citep{Raghunathan2024}, Simons Observatory~\citep{Bhimani2024}, and CMB-S4~\citep{Alvarez2021} will improve constraints on $\tau_{\rm es}$ and pkSZ.  They will may also detect new signals that probe reionization, such as patchy $\tau$~\citep{Coulton2024} and higher-order statistics and cross-correlations with other signals~\citep[e.g.][]{LaPlante2020}.  These will help constrain the early stages of reionization because of their sensitivity to its duration and morphology~\citep{Chen2022}.  

\section{Conclusions} \label{sec:conc}

In this work, we have studied the observational properties of three representative reionization histories. In the first, reionization starts early and ends at $z \sim 8$, earlier than suggested by the Ly$\alpha$ forest and $\tau_{\rm es}$.   This scenario is motivated by recent JWST observations of the UVLF and $\xi_{\rm ion}$ at $z > 6$, which, when combined with observationally motivated assumptions about $f_{\rm esc}$, suggest copious ionizing photon output by high-redshift galaxies.  We have investigated the observational properties of this model, alongside two others in which reionization ends much later at $z \sim 5$, in agreement with the Ly$\alpha$ forest and $\tau_{\rm es}$.  One model starts reionization relatively late at $z \sim 9$, and the other starts early at $z \sim 13$.  

\begin{itemize}

    \item We find, consistent with previous work, that the \textsc{early start/early end} scenario severely violates high-redshift QSO observations, most notably the Ly$\alpha$ forest.  These observations require reionization to end at $5 < z < 6$, or at least not much sooner.  This is consistent with recent measurements of $\tau_{\rm es}$ from~\citealt{Planck2018,deBelsunce2021}.  Unfortunately, neither the mean transmission of the Ly$\alpha$ forest nor $\tau_{\rm es}$ display a clear preference for whether reionization started late or early.  The former measures only the global average transmission of the IGM at reionization's end, and the latter is only an integrated constraint, and thus does not uniquely constrain reionization's early stages.   

    \item Observations of the UVLF and $\xi_{\rm ion}$ by JWST, direct measurements of the MFP from QSO spectra, the visibility of $z \geq 8$ LAEs, and the recent SPT measurement of the patchy kSZ all prefer a late start to reionization.  In light of the latest UVLF measurements at $z \geq 8$ from JWST, our early-starting model requires an order of magnitude of evolution in galaxy ionizing properties (quantified by $\langle f_{\rm esc} \xi_{\rm ion} \rangle_{L_{\rm UV}}$) between $z = 6$ and $12$.  This is less compatible with observations than the flat evolution in our late-starting model.  Direct measurements of the MFP from QSO spectra also prefer a late start, mainly because of the high neutral fraction needed to match direct measurements from QSO spectra at $z = 6$.  The steep drop-off in LAE visibility at $z > 8$ observed by~\citealt{Tang2024b} is more consistent with a late than an early start.  Finally, the low central value of the SPT pkSZ measurement prefers a late start, and disfavors our early-starting model at almost $2 \sigma$.  
    
    \item By contrast, the distribution of Ly$\alpha$ forest opacities, the thermal history of the IGM, and measurements of $x_{\rm HI}$ at $z < 6.5$ prefer an early start.  The forest $P(< \tau_{\rm eff})$ is too wide for the observations in our \textsc{late start/late end} model, preferring instead the lower neutral fraction in the \textsc{early start/late end} case.  Constraints on $x_{\rm HI}$ at $z \leq 6.5$ from a variety of QSO-based inferences suggest a similar conclusion.  The cooler IGM in the early-starting case at $z \leq 5$ is also in better agreement with observations.  

    \item Our findings suggest that no single probe can conclusively rule out either the \textsc{late start/late end} or \textsc{early start/late end} model in favor of the other.  However, we do find that observations across multiple independent datasets - JWST observations of galaxy properties, LAE detections, the CMB, and QSO absorption spectra - prefer a late start to reionization.  The observables that prefer an early start are all derived from the same type of observations (QSO spectra) and only probe the tail end of reionization.  As such, they are not fully independent probes, and they require a model to derive inferences about reionization's early stages.  As such, we conclude that overall, existing observational data displays a mild preference for the \textbf{\textsc{late start/late end}} scenario.   

    \item The face-value disagreement we find between different probes suggests that (1) present observations, and the models used to interpret them, are insufficiently precise and/or accurate to paint a consensus picture of reionization's early stages, and/or (2) there are systematic effects (in observations and/or theoretical modeling) leading to the appearance of tension.  The second possibility motivates care in interpreting the results of analyses using multiple data sets.  Joint analyses using many observables could lead to artificially tight constraints on the reionization history and other quantities if tensions arising from systematics are not carefully understood.   

\end{itemize}

Forthcoming work, on the observational and theoretical side, should continue working to synergize the information available from many observables.  Our work motivates further efforts targeting the early stages of reionization, which will yield key insights into the evolution of galaxy properties across the reionization epoch and into the cosmic dawn era.  This work is also a cautionary tale that motivates careful understanding of potential systematics in both observations and theoretical modeling.  Such systematics, if not studied carefully, could lead to premature conclusions about reionization.  

\begin{acknowledgements}

CC acknowledges helpful conversations with Seth Cohen, Timothy Carleton, Evan Scannapieco, Frederick Davies, and Kevin Croker, and support from the Beus Center for Cosmic Foundations.  A.D.'s group was supported by grants NSF AST-2045600 and
JWSTAR-02608.001-A.  RAW acknowledges support from NASA JWST Interdisciplinary Scientist grants
NAG5-12460, NNX14AN10G and 80NSSC18K0200 from GSFC. JBM acknowledges support from NSF Grants AST-2307354 and AST-2408637, and thanks the Kavli Institute for Theoretical Physics for their hospitality.  Computations were made possible by NSF ACCESS allocation TG-PHY230158. 

\end{acknowledgements}

\bibliography{references}{}
\bibliographystyle{aasjournal}

\appendix

\section{$P(< \tau_{\rm eff}^{50})$ for shifted reionization histories}
\label{app:Ptau}

\begin{figure*}
    \centering
    \includegraphics[scale=0.165]{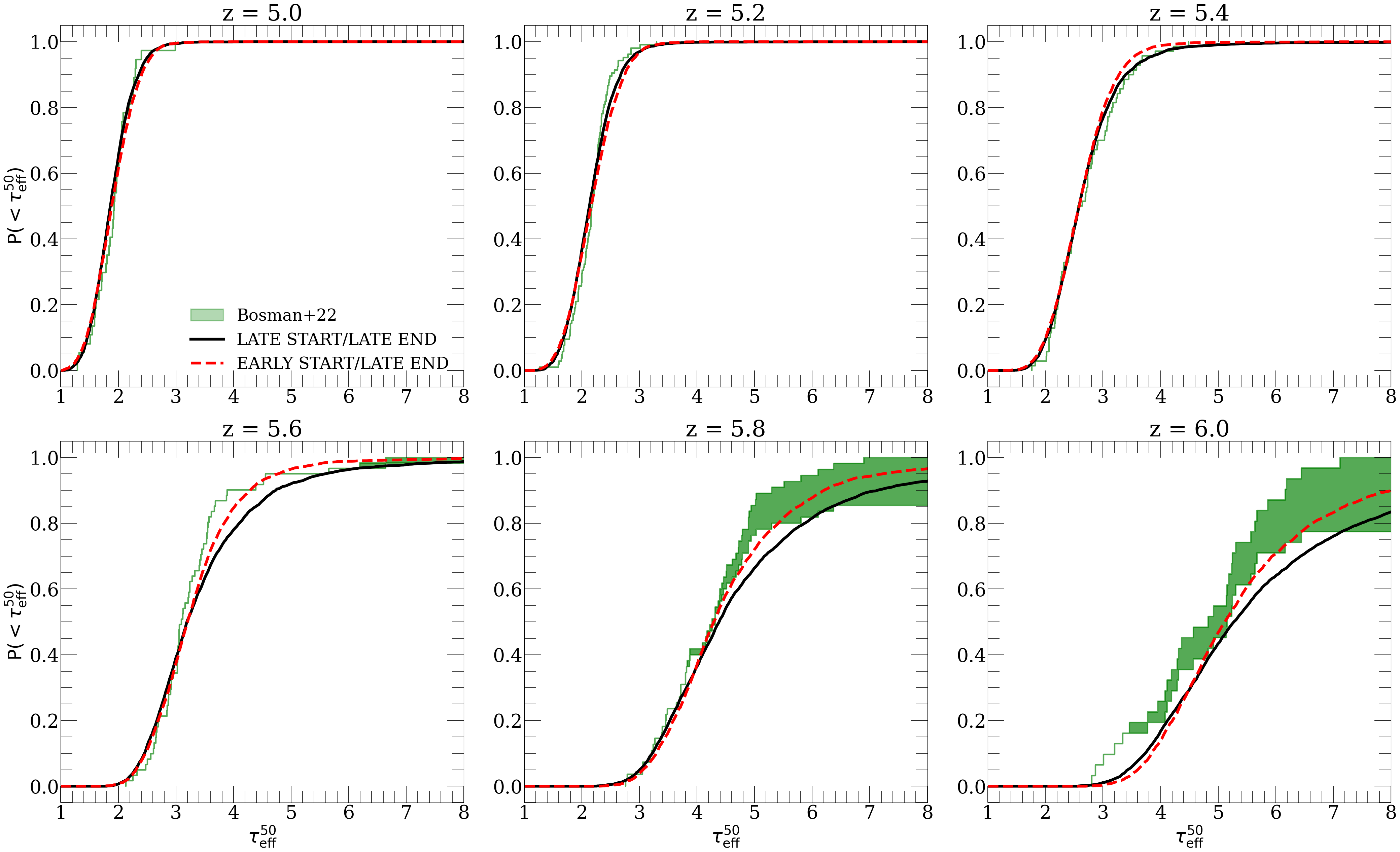}
    \caption{Ly$\alpha$ forest $P(< \tau_{\rm eff}^{50})$ with the reionization histories in our late-ending models shifted earlier by $\Delta z = 0.3$, as described in \S\ref{subsubsec:lya}.  This test estimates what these models would look like if reionization ended at $z = 5.3$ (as suggested by the analysis of~\citealt{Bosman2021}) instead of at $z = 5$.  At $z = 5$, $5.2$, and $5.4$, there is little difference between the two models, as reionization is complete (or nearly so) in both cases.  At $z = 5.6$ and $5.8$, $P(< \tau_{\rm eff}^{50})$ is narrower in the \textsc{early start/late end} model and agrees well with the observations, whereas $P(< \tau_{\rm eff}^{50})$ is still too wide in the \textsc{late start/late end} case.  At $z = 6$, neither of the shifted models agree particularly well with the data, and it is difficult to tell visually which is a better fit.  }
    \label{fig:taueff_appendix}
\end{figure*}

In Figure~\ref{fig:taueff_appendix}, we show $P(< \tau_{\rm eff}^{50})$ for the \textsc{late start/late end} and \textsc{early start/late end} models with their reionization histories shifted earlier by $\Delta z = 0.3$, as described in \S\ref{subsubsec:lya}.  We show $6$ redshifts between $z = 5$ and $6$ in intervals of $0.2$.  This exercise shows roughly what $P(< \tau_{\rm eff}^{50})$ would look like if these models ended reionization at $z = 5.3$ instead of $5$.  We see that at $z = 5$, $5.2$, and $5.4$, there is little difference between the models.  At $z = 5.6$ and $5.8$,  $P(< \tau_{\rm eff}^{50})$ is slightly narrower in the \textsc{early start/late end} case, as it is in Figure~\ref{fig:taueff}.  The difference is that the \textsc{early start/late end} model now seems to agree well with the observations, whereas in the \textsc{late start/late end} case, $P(< \tau_{\rm eff}^{50})$ is still too wide.  At $z = 6$, neither scenario fits the data particularly well, and it is hard to visually assess which is a better fit.  

In Table~\ref{tab:deltaPtau}, we give $\delta P_{\rm data}(<\tau_{\rm eff}^{50})$ for the three models discussed in \S\ref{subsubsec:lya} (top three rows) and for the shifted reionization histories discussed here (bottom rows).  At all redshifts except $z = 6$, the $\Delta z = 0.3$ shift significantly improves the agreement between the data and the \textsc{late start/late end} model.  The agreement with the \textsc{early start/late end} model gets worse at $z = 6$ and $5.4$, significantly better at $z = 5.8$, $5.6$, and $5$, and changes very little at $z = 5.2$.  However, the shifted \textsc{early start/late end} model still agrees with the data better than or as well as the \textsc{late start/late end} case at all redshifts except $z = 5.4$.  This shows that even if reionization ends significantly earlier than it does in our models, the \textsc{early start/late end} scenario remains preferred overall by $P(<\tau_{\rm eff}^{50})$, albeit by a narrower margin than in our fiducial scenarios.

\begin{table}
\begin{tabular}{||c||c|c|c|c|c|c||}
    \hline \hline
    Redshift &  6.0 & 5.8 & 5.6 & 5.4 & 5.2 & 5.0\\
    \hline \hline
    \textsc{late start/late end} & 0.1796 & 0.318 & 1.314 & 0.280 & 0.865 & 0.380 \\
    \textsc{early start/late end} & 0.0438 &  0.270 & 0.589 & 0.0786 & 0.4746 & 0.487 \\
    \textsc{Reionization over by $z > 6$} & 0.275 & 0.240 & 0.0565 & 0.385 & 0.339 & 0.237 \\
    \textsc{late start/late end (shifted)} & 0.234 & 0.199 & 0.313 & 0.0893 & 0.371 & 0.149 \\
    \textsc{early start/late end (shifted)} & 0.148 & 0.0798 & 0.164 & 0.1612 & 0.482 & 0.110\\
     \hline \hline
\end{tabular}
\label{tab:deltaPtau}

\caption{$\delta P(<\tau_{\rm eff}^{50})$ at all $6$ redshifts between $z = 6$ and $5$, spaced by $\Delta z = 0.2$, for each of our scenarios.  The top three rows give values for the models described in the main text, and the bottom two for the shifted models shown here.  In all cases except $z = 6$, the goodness of fit for the \textsc{late start/late end} model improves significantly when it is shifted earlier by $\Delta z = 0.3$.  The fit for the \textsc{early start/late end} model gets worse at $z = 6$ and $5.4$, improves at $z = 5.8$, $5.6$, and $5$, and changes little at $z = 5.2$.  At $z = 6$, $5.8$, and $5.6$, the fit is still notably better for the shifted \textsc{early start/late end} model than for the \textsc{late start/late end} case.  }
\end{table}

\section{Ly$\alpha$ visibility in the full $(T_{\rm thresh}, v_{\rm off})$ parameter space}

\begin{figure*}
    \centering
    \includegraphics[scale=0.145]{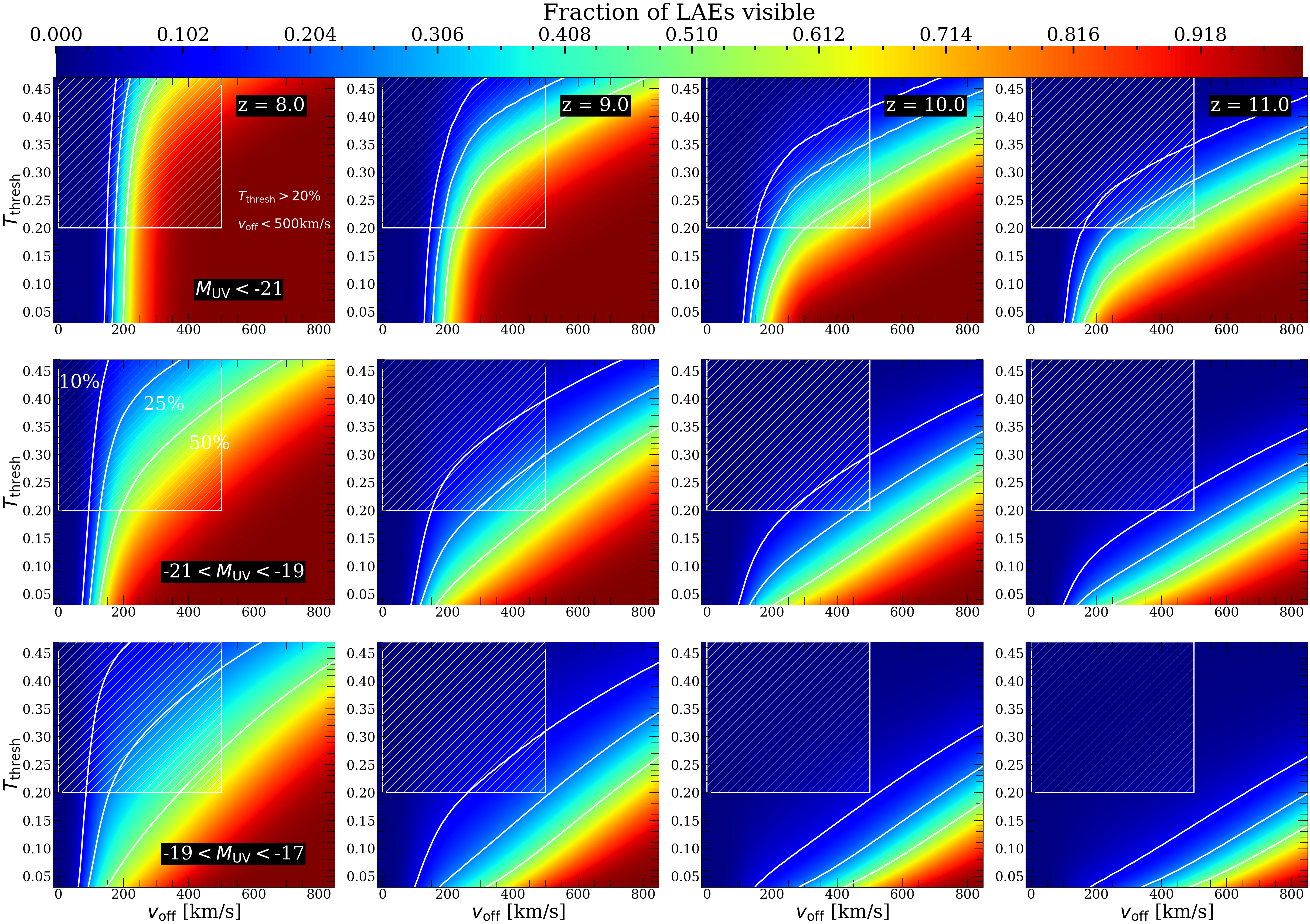}
    \caption{Summary of LAE transmission statistics in the \textsc{late start/late end} model.  The y axis is $T_{\rm thresh}$ (Eq.~\ref{eq:Tthresh}) and the x axis is the $v_{\rm off}$ at line center.  The color plot shows the fraction of visible LAEs (with $T_{\rm IGM} > T_{\rm thresh}$) across this parameter space.  The panels denote different redshifts (columns) and ranges of $M_{\rm UV}$ (rows).  The white contour lines denote ``visibility fractions'' of $10\%$, $25\%$, and $50\%$.  The hatched region in the upper left of each panel denotes $T_{\rm thresh} > 0.2$ and $v_{\rm off} < 500$ km/s.  }
    \label{fig:transmission_fraction}
\end{figure*}

In this appendix, we show complete results for our LAE visibility analysis in terms of $v_{\rm off}$ and $T_{\rm thresh}$ presented in \S\ref{subsubsec:LAE_visibility}.  Figure~\ref{fig:transmission_fraction} shows the fraction of visible LAEs at $T_{\rm thresh} < 0.5$ and $0 < v_{\rm off} < 800$ km/s, at the four redshifts (columns) and three magnitude ranges (rows) we considered in the \textsc{late start/late end} model.  Red (blue) regions denote high (low) LAE visibility fractions.  The white contour lines denote visibility fractions of $50\%$, $25\%$, and $10\%$ (see annotation in the left-most panel of the middle row).  The hatched white box in the upper left corner of each panel denotes the region where $T_{\rm thresh} > 0.2$ and $v_{\rm off} < 500$ km/s.  We expect a majority of $z \geq 8$ LAEs to inhabit this region.  Most LAEs in $M_{\rm UV} < -17$ galaxies observed at slightly lower redshifts ($z \sim 5-6$, when $T_{\rm IGM}$ is close to unity) have $v_{\rm off} < 500$ km/s and EW$_{\rm int} < 500$ $\text{\AA}$~\citep[][]{Goovaerts2023,Tang2024a}.  For $T_{\rm thresh} = 0.2$, the latter would correspond to EW$_{\rm obs}^{\min} < 100$ $\text{\AA}$.  

At $z = 8$, a significant portion of the hatched region displays a high visibility fraction, especially for the brightest galaxies.  For $M_{\rm UV} < -21$ galaxies, LAEs with $v_{\rm off} > 200$ km/s have a significant chance of being visible, provided they are bright enough to be observed with a factor of $2$ IGM attenuation.  Fainter galaxies require somewhat fainter detection thresholds, since on average they inhabit smaller ionized bubbles and are more sensitive to IGM attenuation.  In the faintest $M_{\rm UV}$ bin, $v_{\rm off} > 400$ kms/ and $T_{\rm thresh} < 0.25$ is required for half of LAEs to be visible.  Still, we should expect to observe some LAEs at $z = 8$ in the \textsc{late start/late end} model, especially the brightest ones.  

At $z > 8$, the transmission of Ly$\alpha$ declines rapidly, especially for fainter galaxies.  At these redshifts, in all but the brightest $M_{\rm UV}$ bin, the $50\%$ contour line does not intersect the hatched region.  As we showed in Figure~\ref{fig:transmission_evolution}, this drop-off in visibility is consistent with the observed decline in $X_{\rm Ly\alpha}$ observed by~\citealt{Tang2024b} at $z > 8$.  The brightest objects remain likely to be observed if $v_{\rm off} > 400$ km/s and $T_{\rm thresh} < 0.25$ all the way to $z = 11$.  Notably, GN-z11 ($M_{\rm UV} = -21.5$, $z = 10.6$,~\citealt{Bunker2023}) has $v_{\rm off} = 550$ km/s, meeting this condition.  The recently observed JADES-GS-z9-0 ($M_{\rm UV} = -20.43$, $z = 9.4$,~\citealt{Curti2024}) has $v_{\rm off} = 450$ km/s.  These objects would be likely visible in the \textsc{late start/late end} model if their intrinsic EWs were in the neighborhood of $100$ $\text{\AA}$, $\approx 4\times$ larger than their observed EWs (see \S\ref{subsec:gnz11}).  These results are consistent with a universe in which most LAEs at $z > 8$ are obscured by the IGM, but a small number of objects that are relatively bright, have high $v_{\rm off}$, and/or high intrinsic EWs remain visible.    

\begin{figure*}
    \centering
    \includegraphics[scale=0.145]{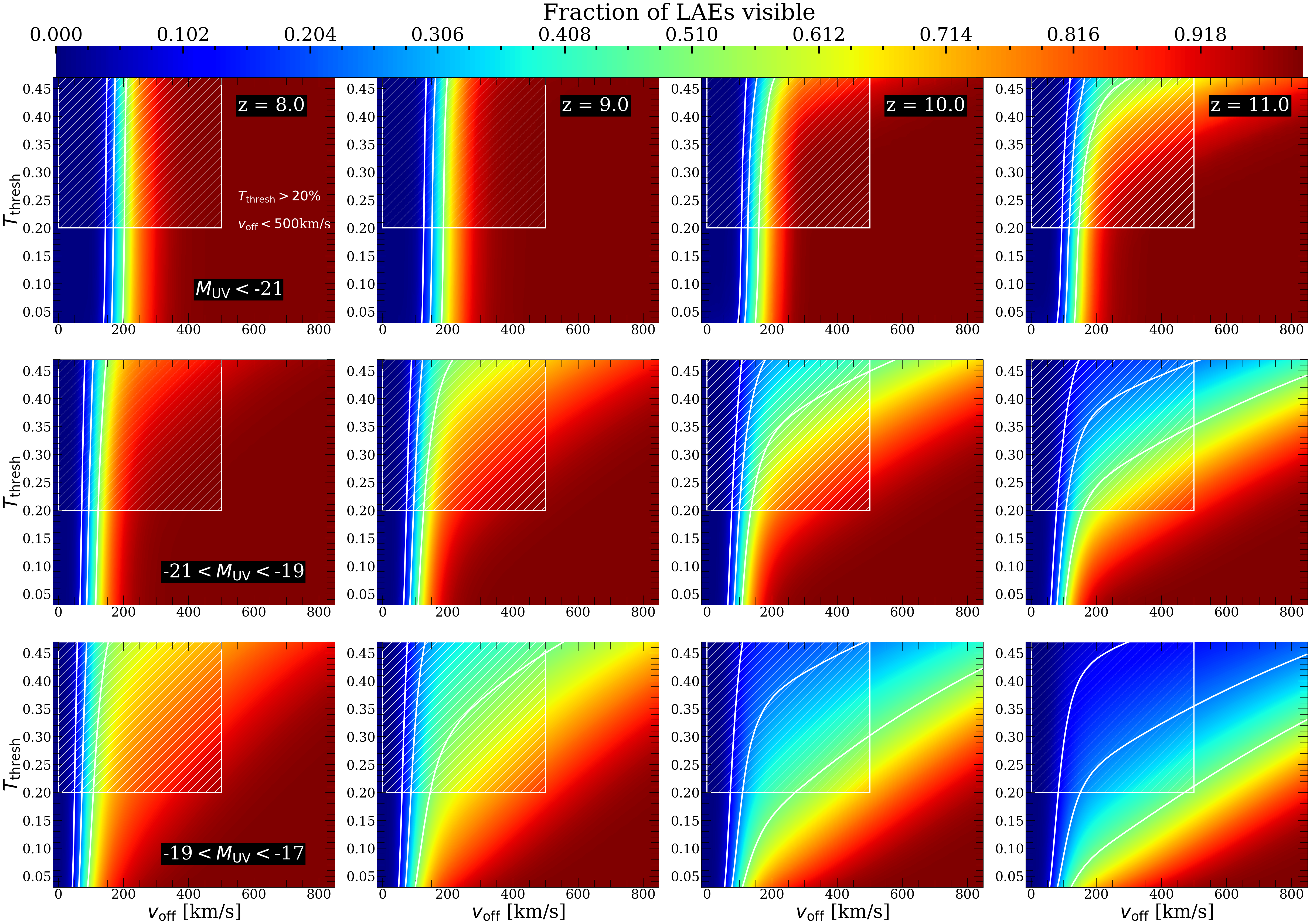}
    \caption{Same as Figure~\ref{fig:transmission_fraction}, but for the \textsc{early start/late end} model.  A significant fraction of LAEs with typical properties (even faint ones) should remain visible in this scenario even at $z = 11$, in contrast with the observations~\citealt{Tang2024b}.  }
    \label{fig:transmission_fraction_early}
\end{figure*}

Figure~\ref{fig:transmission_fraction_early} is the same as Figure~\ref{fig:transmission_fraction}, but for the \textsc{early start/late end} model.  In contrast to the \textsc{late start/late end} case, a significant fraction of the parameter space displays high visibility, even at $z = 11$.  Objects with $M_{\rm UV} < -21$ are likely to be visible up to $z = 11$ as long as $T_{\rm thresh} \leq 0.5$ and $v_{\rm off} > 200$ km/s.  Even the faintest galaxies have a significant chance of being observed at $z = 11$.  It should then be expected that in such a scenario, a significant fraction of LAEs - even faint ones - with typical physical properties should remain visible up to $z = 11$.  At face value, the observed sharp decline in LAE visibility across the population of LAEs up to this redshift does not prefer this scenario.  

\end{document}